\documentclass[letterpaper,twocolumn,10pt]{article}
\usepackage{usenix2019_v3}

\usepackage{tikz}
\usepackage{amsmath}

\usepackage{amsmath,amssymb,amsfonts}
\usepackage{amsmath}

\usepackage{tabularx}
\usepackage{booktabs}

\usepackage{multirow}

\usepackage{pifont}
\newcommand{\xmark}{\ding{55}}%

\usepackage{algorithm}
\usepackage[noend]{algpseudocode}

\algtext*{EndWhile}% Remove "end while" text
\algtext*{EndIf}% Remove "end if" text
\algtext*{EndFor}% Remove "end if" text

\usepackage{graphicx}

\usepackage{textcomp}
\usepackage{xcolor}

\usepackage{subcaption} 

\usepackage{filecontents}

\begin{document}
%-------------------------------------------------------------------------------

%don't want date printed
\date{}

\title{COMA: Communication and Obfuscation Management Architecture }

%for single author (just remove % characters)
\author{
{\rm Kimia Zamiri Azar\IEEEauthorrefmark{1}, Farnoud Farahmand\IEEEauthorrefmark{1}, Hadi Mardani Kamali\IEEEauthorrefmark{1}, Shervin Roshanisefat\IEEEauthorrefmark{1},}\\
\and
{\rm Houman Homayoun\IEEEauthorrefmark{3}, William Diehl\IEEEauthorrefmark{2}, Kris Gaj\IEEEauthorrefmark{1}, Avesta Sasan\IEEEauthorrefmark{1}}\\ \\
\IEEEauthorrefmark{1} Department of ECE, George Mason University, VA, USA. \\
\{kzamiria, ffarahma, hmardani, sroshani, kgaj, asasan\}@gmu.edu
% copy the following lines to add more authors
\and
{}
\IEEEauthorrefmark{3} Department of ECE, University of California, Davis, CA, USA. \{hhomayoun\}@ucdavis.edu
\and
{}
\IEEEauthorrefmark{2} Department of ECE, Virginia Tech, VA, USA. \{wdiehl\}@vt.edu
} % end author

\maketitle

%-------------------------------------------------------------------------------
\begin{abstract}
%-------------------------------------------------------------------------------
In this paper, we introduce a novel \emph{Communication and Obfuscation Management Architecture} (COMA) to handle the storage of the obfuscation key and to secure the communication to/from untrusted yet obfuscated circuits. COMA addresses three challenges related to the obfuscated circuits: First, it removes the need for the storage of the \emph{obfuscation unlock key} at the untrusted chip. Second, it implements a mechanism by which the key sent for unlocking an obfuscated circuit changes after each activation (even for the same device), transforming the key into a dynamically changing license. Third, it protects the communication to/from the COMA protected device and additionally introduces two novel mechanisms for the exchange of data to/from COMA protected architectures: (1) a highly secure but slow double encryption, which is used for exchange of key and sensitive data (2) a high-performance and low-energy yet leaky encryption, secured by means of frequent key renewal. We demonstrate that compared to state-of-the-art key management architectures, COMA reduces the area overhead by 14\%, while allowing additional features including unique chip authentication, enabling activation as a service (for IoT devices), reducing the  side channel threats on key management architecture, and providing two new means of secure communication to/from an untrusted chip.

\end{abstract}

\section{Introduction} 
\vspace{-2mm}
The increasing cost of IC manufacturing has pushed several stages of the semiconductor device's manufacturing supply chain to the third-party facilities, which are identified as untrusted entities \cite{yeh2012trends}. Fabrication of ICs in an untrusted supply chain has introduced multiple forms of security threats such as the possibility of overproduction, Trojan insertion, Reverse Engineering (RE), Intellectual Property (IP) theft, and counterfeiting \cite{rostami2014primer, tehranipoor2017invasion}. The stage that poses the utmost vulnerability is the fabrication stage, in which an untrusted foundry has the ultimate knowledge about a to-be-fabricated IC, and with minimal effort could reverse engineer the \emph{GDSII} to its gate-level netlist, analyze, copy, and/or alter the design, creating trust and security challenges for the original design house. 

Considering that a foundry has the ultimate knowledge about the design, passive protection techniques such as watermarking, IC metering, or camouflaging  \cite{alkabani2007active, kahng1998watermarking, cocchi2013method, roy2010ending} are not well suited to protect against attacks initiated at this stage of supply chain, although they can be used to either identify counterfeits, or prevent reverse engineering of the manufactured ICs post fabrication. To protect the IP from being reverse engineered, overproduced, or stolen in the manufacturing supply chain, researchers have studied various means of hardware obfuscation \cite{roy2010ending, Kamali:2019:FHD:3316781.3317831, Kolhe:2019:CLO:3299874.3319496, 8429401, Roshanisefat:2018:SSC:3194554.3194596 ,7951830, kolhe2019customlut, yasin2016sarlock, xie2016mitigating, yasin2017provably}, which is the process of hiding the true functionality of an IC when no key, or an incorrect key, is present. Only once the correct key is provided, the IC behaves correctly. The requirement for obfuscated solutions is to resist various forms of attacks against such circuits including brute force, sensitization, Boolean satisfiability (SAT) or satisfiability modulo theories (SMT), removal, approximate-based, signal probability skew, functional analysis, etc. \cite{rajendran2013security, subramanyan2015evaluating, azar2019smt, yasin2017security, sirone2018functional, ZamiriAzar:2019:TLL:3299874.3319495, 8474189, sail_bhuni}. 

To remain hidden, in addition to resisting the attacks against its obfuscated circuit(s), the IC should also resist passive, active, or destructive attacks that could be deployed to read the key values. Hence, neither the activation of such devices nor the storage of key values in them should expose or leak the key information. Activation of an obfuscated IC requires storing the activation key in a secure and tamper-proof memory. \cite{tuyls2006read, guajardo2007physical}. However, there exist a group of applications that could use an alternative key storage. This alternative solution is to store the key outside the IC, where the IC is activated every time it is needed. This option requires constant connectivity to the key management source and a secure communication for key exchange to prevent any leakage of the key. This solution allows an IC designer to store the chip unlock key outside of an untrusted chip. So, no secure and tamper-proof memory is needed. Since the key is stored outside the untrusted chip, a constant connectivity to an obfuscation key-management solution is an indispensable requirement for this category of devices. This requirement could be easily met for two prevalent groups of architectures: (1) 2.5D package-stack devices where a single trusted chip is used for key management and activation of multiple obfuscated ICs manufactured in untrusted foundries, and (2) IoT devices with constant connectivity to the cloud/internet.

In 2.5D package-integrated ICs, similar to DARPA SPADE architecture \cite{darpaspade}, a chip which is fabricated in a trusted foundry, but in a larger technology node, is packaged with an untrusted chip fabricated in an untrusted foundry in a smaller technology node. The resulting solution benefits from the best features of both technologies: The untrusted chip may be used as an accelerator, providing the resulting hybrid solution with the much-needed scalability (higher speed and lower power), while the trusted chip provides the means of trust and security. The untrusted chip is isolated from the outside world and any exchange of information to/from untrusted chip passes through the trusted chip. 

The second group of devices in this category are IoT devices, where constant connectivity is their characterizing features. In these solutions the obfuscation key could be stored in the cloud, and activation of an IoT device could be done remotely. This model allows custom, monitored, and service oriented activation (Activation As A Service). An additional advantage is removing the possibility of extracting an unlock key from a non-volatile memory that otherwise would have to be used for storing the obfuscation unlock key. Examples of which are IoT devices used for providing various services, military drones activated for a specific mission, video decryption services for paid pay-per-view transactions, etc., where a device has to operate in an unsafe environment and is at risk of capture and reverse engineering. In these applications, the IC fabricated in an untrusted foundry is activated either every time it is powered up, or for certain time intervals. The key vanishes after the service is performed, or when the device is powered down. The activation of such devices is performed using a remote key management service (in the cloud or at a trusted base-station), and the key exchange to/from these devices should be secured.

In both 2.5D system solutions and IoT devices, the need for implementation of a tamper-proof memory, for storage of IC activation key, in an untrusted process is removed. Some reasons why implementing a secure memory in an untrusted foundry may be undesired, or practically unfeasible include:

\textit{Availability:} The targeted foundry may not offer the required process for implementing a secure memory with the desired features. An example could be the requirement for storing sensitive information in magnetic tunnel junction (MTJ) memories to prevent probing and attacks that could be deployed against flash-based NVMs. Fabricating such ICs requires a hybrid process that supports both CMOS and MTJ devices, which may be unsupported by the targeted foundry. 

\textit{Verified Security:} The secure memory may be available in the targeted technology, however not be fully tested and verified in terms of its resistance against different attacks.

\textit{Cost:} Implementing secure memory requires additional fabrication layers and processing steps, increasing the cost of manufacturing. Increasing the silicon area is a far cheaper solution than increasing the number of fabrication layers.

\textit{Reusability:} In 2.5D system solutions, a trusted chip could be shared by multiple untrusted chips, manufactured in different foundries. Moving the secure memory to the trusted chip removes the need for implementing the secure memory in all utilized processes. The trusted chip could be designed once with utmost security for protection and integrity of data. This also reduce the cost of manufacturing untrusted chips by removing the need for additional processing steps for implementing secure memory.  

\textit{Ease of Design:} Implementing secure memory requires pushing the design through non-standard physical design flow to implement the tamper-proof layers in silicon and package. In addition, the non-volatile nature of tamper-proof memory requires read and write at elevated voltages, increasing the burden on the power-delivery network design. Reuse of a trusted chip with a tamper proof memory that could manage activation of other obfuscated ICs, relaxes the design requirement of ICs to standard physical design and fabrication process.

% \textcolor{red}{\textit{Hard Design Requirement:} The requirement for the memory features and computational requirements (in terms of PPA) may not be satisfied in the same process. For example, when for a mission-critical system a radiation hardened memory is required, the superior choice is a Si-Ge process, where the Power, Performance and Area (PPA) could be best addressed in an advanced CMOS process.}

% \textit{Availability of Advanced Tamper-Resistant Features:} Certain foundries may specialize in providing advanced security features that are not available in the standard processing flows. An example is a process that supports chemical (acid) vanishability, where upon tamper detection, a chemical reaction dissolves the netlist.

In this paper, we propose the COMA key-management and communication architecture for secure activation of obfuscated circuits that are manufactured in untrusted foundries and meet the constant connectivity requirement, namely ICs that belong to a) 2.5 package-integrated and b) IoT solutions. We describe two variants of our proposed solutions: The first variant of COMA is used for secure activation of IPs within 2.5D package-integrated devices (similar to DARPA SPADE). The second variant of COMA is used for secure activation of connected IoT devices. The proposed COMA allows us to (1) push the obfuscation key and obfuscation unlock mechanism off of an untrusted chip, (2) make the key a moving target by changing it for each unlock attempt, (3) uniquely identify each IC, (4) remove the need to implementing a secure memory in an untrusted foundry, and (5) utilize two novel mechanisms for ultra-secure or ultra-fast encrypted communication.

The rest of this paper is organized as follows: Section~\ref{background} presents the background and related work to secure key-exchange and obfuscation schemes. Section~\ref{proposed} demonstrates how the proposed method has significant advantages in terms of security and performance. Both variants of the proposed architecture are evaluated in this Section. The security of the proposed architecture against various attacks is discussed in Section \ref{attacks}. The experimental results, as well as comparison with prior-art methods, is presented in Sections \ref{results} and \ref{comparison_with_prior_work}. Finally, Section \ref{conclusion} concludes the paper.

%(i.e. in the 2.5D stacked architecture and IoT devices) 

%%%\vspace{-2mm}
\section{Background} \label{background}
\vspace{-2mm}

Active metering, Secure Split-Test, logic obfuscation, and solutions such as Ending Piracy of Integrated Circuits (EPIC) have been proposed to protect ICs from supply chain-related security threats by initializing the HW control to a locked state at power-up and hiding the design intent \cite{alkabani2007active, roy2010ending, chakraborty2009harpoon, wendt2014hardware, koushanfar2012provably, baumgarten2010preventing, rajendran2012security, contreras2013secure}. Some of these techniques support single activation, while others support active metering mechanisms. Active metering techniques \cite{alkabani2007active, chakraborty2009harpoon, koushanfar2012provably, rajendran2012security} provide a mechanism for the IP owner to lock or unlock the IC remotely. In these solutions, the locking mechanism is a function of a unique ID generated for each IC, possibly and preferably by a \emph{Physical Unclonable Function} (PUF) \cite{tuyls2006read}. Only the IP owner knows the transition table and can unlock the IC. Active metering, combined with a PUF, makes the key a moving target from chip to chip. However, there exist a few issues with previous metering techniques: first, the key(s) to unlock each IC remains static. Second, these techniques unlock the chips before they are tested by the foundry. Hence, the IP owner can control how many ICs enter the supply chain, but not how many properly tested ICs exit the supply chain. Finally, these techniques do not respond well to the threat of the foundry requesting more IDs by falsifying the yield to be lower during the test process. Such shortcomings can potentially allow the foundry to ship more out-of-spec or defective ICs to the supply chain. 

\begin{figure}[t]
\centering
\includegraphics[width = \columnwidth]{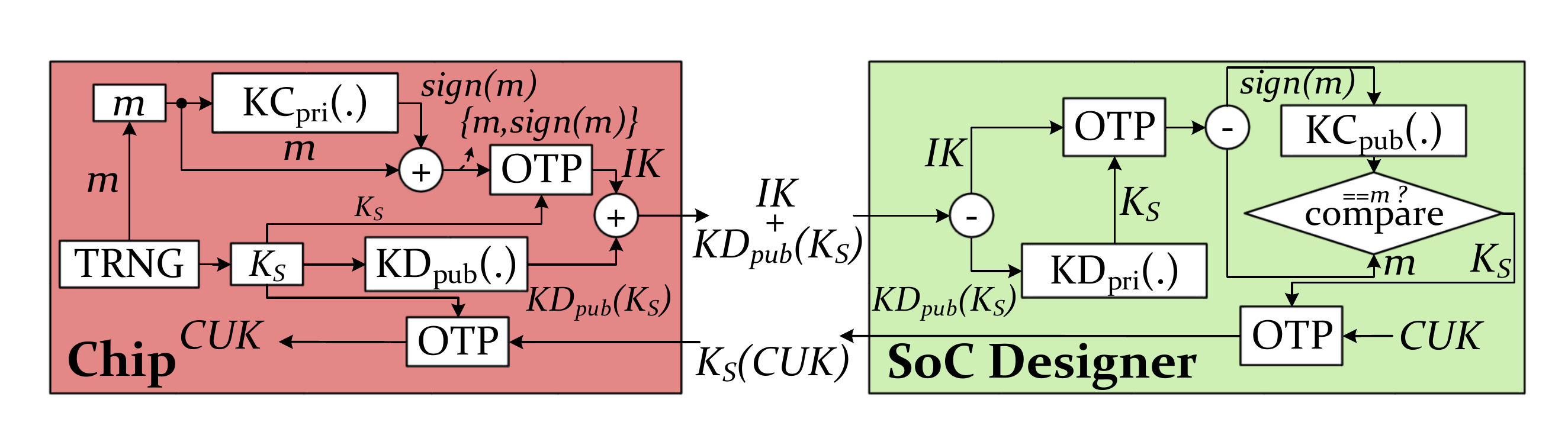}
% %%\vspace{-5mm}
\caption{FORTIS: Overall Architecture. 
% In this figure OTP is One Time Pad, CUK is the chip unclock key, $C_{pri}$ is private key cryptography, $KD_{pub}$ is the public key cryptography, and TRNG is the true random number generators. 
}
\label{fortis_arch}
%%\vspace{-4.5mm}
\end{figure}

Many of these shortcoming were addressed in FORTIS \cite{guin2016fortis} shown in Fig. \ref{fortis_arch}. In FORTIS the registers that hold the obfuscation key are made a part of the scan chain, allowing the foundry to carry structural test by assigning test values to these registers prior to the activation of the IC. Authors of \cite{guin2016fortis} argue that placing a DFT compression logic, not only reduces the test size, but also prevents the readout of the individual register values. After testing the IC, the obfuscation key is transferred and applied to unlock the circuit using two types of cryptographic modules: a public-key crypto engine, and a One Time Pad (OTP) crypto engine. 

%%\vspace{2pt}
In FORTIS, the public and private keys are hardwired in the design. A TRNG is used to generate a random number (m) that is treated as a message. This message is encrypted using the private key of the chip to generate a signature sig(m). The actual message and its signature are concatenated and later used as a mean for the authentication of the chip. At the same time, the TRNG generates another random number $K_S$. This random number is used as the key for OTP, and at the same time is encrypted using the public key of the designer to generate $KD_{pub}(K_S)$. OTP uses $K_S$ for encrypting the ($m,sig(m)$), and the output of OTP is concatenated with the $KD_{pub}(K_S)$. The resulting string of bits is transmitted to the SoC designer. The SoC designer uses a OTP to obtain $m$ and $sig(m)$ for the purpose of authentication. She then uses the private key of the designer to recover $K_S$. Finally, $K_S$ is used by OTP to encrypt the chip unlock key (CUK). The encrypted CUK is transmitted to the chip, decrypted using OTP, and applied to the obfuscation unlock key registers to unlock the circuit.

% \begin{figure}
%     \centering
%     \begin{subfigure}[b]{\columnwidth}
%     \includegraphics[width=\textwidth]{images/fortis.pdf}
%     \caption{FORTIS: Overall Architecture~\cite{guin2016fortis} %%\vspace{4pt}}
%   \end{subfigure}
%   \begin{subfigure}[b]{\columnwidth}
%     \includegraphics[width=\columnwidth]{images/comavsfortis.pdf}
%     \caption{COMA: Overall Architecture %%\vspace{-5pt}}
%   \end{subfigure}
%     \caption{The Key Management Architecture of a) FORTIS (prior work) and b) COMA (this work) %%\vspace{-10pt}}
%     \label{fortis_arch}
% \end{figure}

FORTIS, however, suffers from several security issues including 1) using identical public and private keys in all manufactured chips, and thus its inability for unique device authentication, 2) being vulnerable to modeling attack in which the FORTIS structure is modeled in software for requesting the CUK from SoC designer 3) being vulnerable to side channel attacks on public-key encryption engine aimed at recovering the private key of the chip, 4) being vulnerable to fault attacks in which the value of $K_S$ is fixated, 5) requiring a secure memory for storage of the obfuscation unlock key, and 6) not addressing the mechanism for generating a unique and truly random seed to initialize PRNG. After describing our proposed solution, in section \ref{comparison_with_prior_work}, we explain how these vulnerabilities are addressed in our proposed solution. 

Our proposed solution fits the category of active metering techniques. The key is neither static nor stored in the untrusted chip. A key that is used to activate the IC at the test time cannot be reused to activate the same or a different IC in the future. Hence, the test facility is able to accomplish the test process using ATPG tools with a key which is valid for structural/functional test and it is not valid for any subsequent activation. Additionally, the communication to/from IC is secured using a side-channel protected cryptographic engine, combined with a dynamic switching and inversion structure that enhances the security of the chip against invasive and side-channel attacks. We demonstrate that COMA provides two useful means of secure communication to/from the untrusted chip, one for added security, and one for supporting a higher throughput. The proposed architecture is a comprehensive solution for the key management of the obfuscated IPs, where the challenges related to the activation of the IC and secure communication to/from the IC are addressed at the same time. However, as discussed earlier, it is not a universal solution and would fit within the context of IoT-based solutions or within 2.5D package-integrated solutions, as this solution requires constant connectivity. 

%\vspace{-2mm}
\section{Proposed COMA Architecture} \label{proposed}
\vspace{-2mm}

\begin{figure}[t]
\centering
\includegraphics[width = \columnwidth]{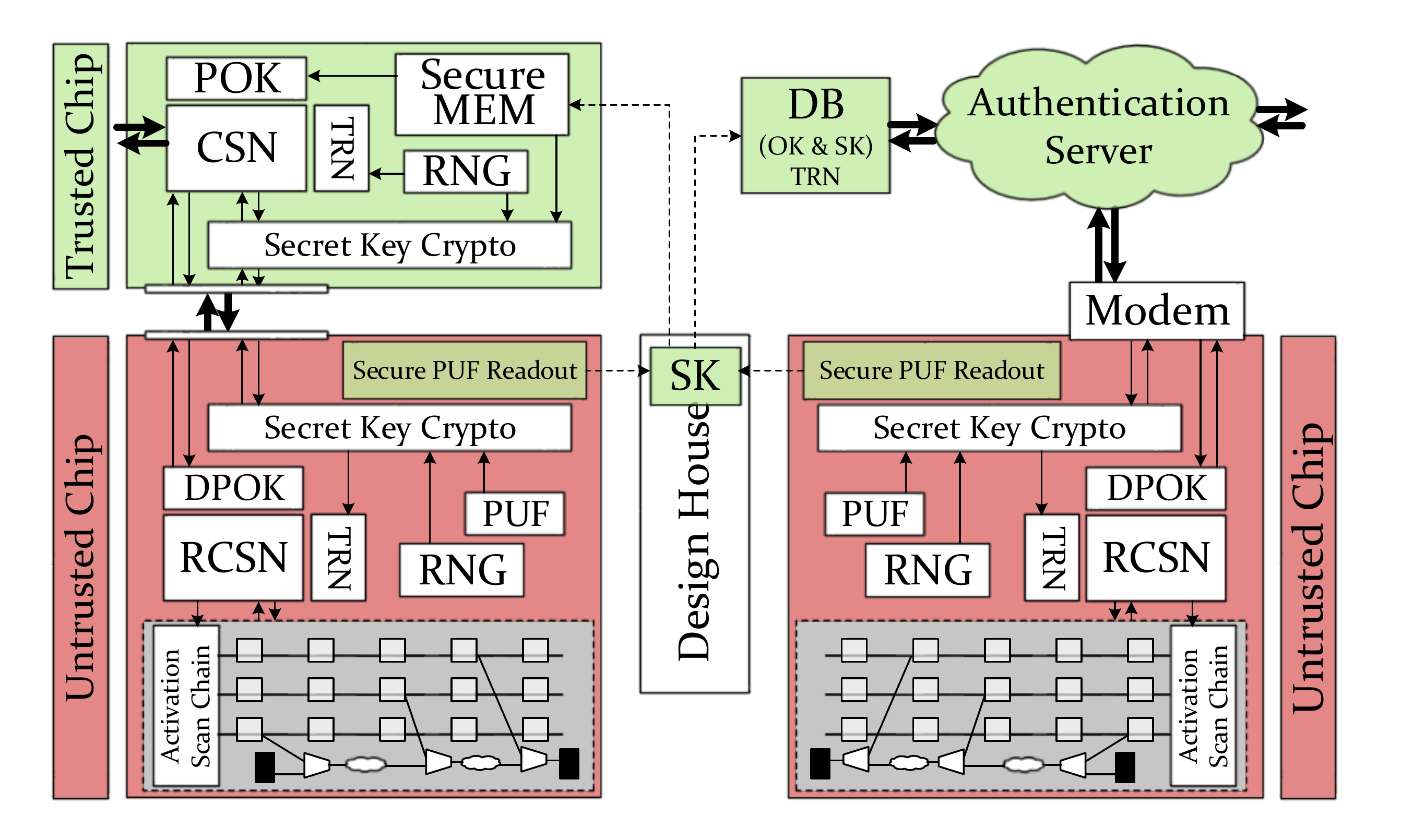}
\caption{Proposed COMAs for (left) 2.5D and (right) IoT-based/remote devices.\vspace{-3mm}}
%\vspace{-4.5mm}
\label{COMA_arch}
\end{figure}

The primary goal of the COMA is to remove the need for storing the \emph{obfuscation key} (OK) on an untrusted chip while securing the communication flow used for activation of the obfuscated circuit in the untrusted chip. The additional benefits of the proposed architecture are the implementation of two new modes of 1) highly secure and 2) very high-speed encrypted communication. We propose two variants of the COMA architecture: The first variant is designed for securing the activation of the obfuscated IP and communication to/from an untrusted IC in 2.5D package-integrated architectures similar to the DARPA SPADE architecture \cite{darpaspade} (denoted by 2.5D-COMA). The second proposed architecture is designed for protecting IoT-based or remotely activated/metered devices (denoted by R-COMA). Fig. \ref{COMA_arch} captures the overall architecture of two variants of the proposed COMAs.

%\vspace{-2mm}
\subsection{2.5D-COMA:  Protecting 2.5D package integrated system solutions}
\vspace{-2mm}

The DARPA SPADE project \cite{darpaspade} explores solutions in which an overall system is split-manufactured between two different technologies, In this solution, a trusted IC which is constructed in an older yet secure technology is packaged with an IC fabricated in an untrusted foundry in an advanced geometry. The purpose of this solution is to provide the best of two worlds: the security of older yet trusted technology and the scalability, power, and speed of the newer yet untrusted technology. 
The 2.5D-COMA is designed to work with an architecture similar to the DARPA SPADE architecture. The proposed solution allows an entire or partial IP in an untrusted chip to be obfuscated, while pushing the mechanism for unlocking and secure activation of the untrusted chip out to a trusted chip. In this solution, the trusted chip encapsulates the sensitive information, verifies the integrity of the untrusted chip, performs sensitive logic monitoring, and controls the activation of the untrusted chip. 
%In this architecture, 
Also, the key to unlock the obfuscated circuit changes per activation, details of which will be explained shortly. 

As shown in Fig. \ref{COMA_arch}, the two variants of COMA contain two main parts, the trusted side (\emph{green}) and the untrusted side (\emph{red}). In both variants, the architectures of untrusted chips are identical, and only the architectures of trusted sides are different. In 2.5D-COMA, only the trusted chip is equipped with a secure memory. The secure memory stores the \emph{Obfuscation Key} (OK) and the Secret Key (SK) used for encrypted communication between the trusted and untrusted chips. The SK is generated using a PUF in the untrusted chip, thus it is unique for each untrusted chip, and the untrusted chip does not need a secure memory to store the SK. The \emph{Configurable Switching Network} (CSN) and \emph{Reverse} CSN (RCSN) are logarithmic routing and switching networks. They are capable of permuting the order and possibly inverting the logic levels of their primary inputs while these signals are being routed to different primary outputs. The RCSN is the exact inverse of the CSN. Hence, passing a signal through CSN-RCSN (or RCSN-CSN) will recover the original input. The switching and inversion behavior of CSN-RCSN is configured using a \emph{True Random Number} (TRN). This TRN is generated in the trusted chip to avoid any potential weakening/manipulating of the TRNG. In addition, since the TRNG in COMA is equipped with standard-statistical-tests applied post-fabrication, such as Repetition-Count test and the Adaptive-Proportion test, as described in NIST SP 800-90B \cite{barker2012recommendation}, any attempt at weakeningthe TRNG during regular operation (i.e. fault attack) can be detected by continuously checking the output of a source of entropy for any signs of a significant decrease in entropy, noise source failure, and hardware failure. By using TRN for the CSN-RCSN configuration, any signal passing through the CSN is randomized, and then by passing through the RCSN is recovered. Additional details are provided in section \ref{CSN_section}.

% \begin{figure}
%     \centering
%     \includegraphics[width = \columnwidth]{images/coma_flow.pdf}  
%     %\vspace{-10pt} 
%     \caption{Unlocking the Untrusted Chip in COMA Architecture. %\vspace{-20pt}}
%     \label{unlock_process}
% \end{figure}

The untrusted chip unlock process in COMA is as follows:
 Prior to each activation, the CSN and RCSN are configured with the same TRN. Since the SK is a PUF-based key generated at the untrusted side, first the SK must be securely readout from untrusted chip. This is done by deploying public key cryptography, the details of which are described in section \ref{secure_puf_readout}. Then, the trusted chip encrypts the TRN using the SK and sends it to the untrusted chip. To perform an activation, as shown in Fig. \ref{COMA_arch}, the OK is read in segments, denoted as \emph{Partial Obfuscation Key} (POK), and is passed through the CSN and encryption on the trusted side and the decryption and RCSN on the untrusted side. This process is repeated every time the obfuscated circuit in the untrusted chip is to be activated, each time using a different TRN for configuring the CSN-RCSN. Usage of a different TRN as the configuration input for the CSN-RCSN for each activation randomizes the input data to Secret key crypto engine. Hence, by using a different TRN for each activation, the obfuscation key (after passing through CSN) is transformed into a one-time license, denoted as \emph{Dynamic Activation License} (DAL). Since the OK is read and sent in segments (from trusted chip), the DAL will be received (at untrusted chip) in segments, denoted as \emph{Dynamic Partial Obfuscation Key (DPOK)}, shown in Fig.~\ref{COMA_arch}, and is used as an input to RCSN. Passing DPOKs through RCSN recovers the POKs, and concatenating the POKs will generate the OK. Note that the DAL is only valid until the TRN is changed. So, the DAL cannot be used to activate other chips or the same chip at a later time. 

\begin{figure}
    \centering
  \begin{subfigure}[b]{\columnwidth}
    \includegraphics[width=\columnwidth]{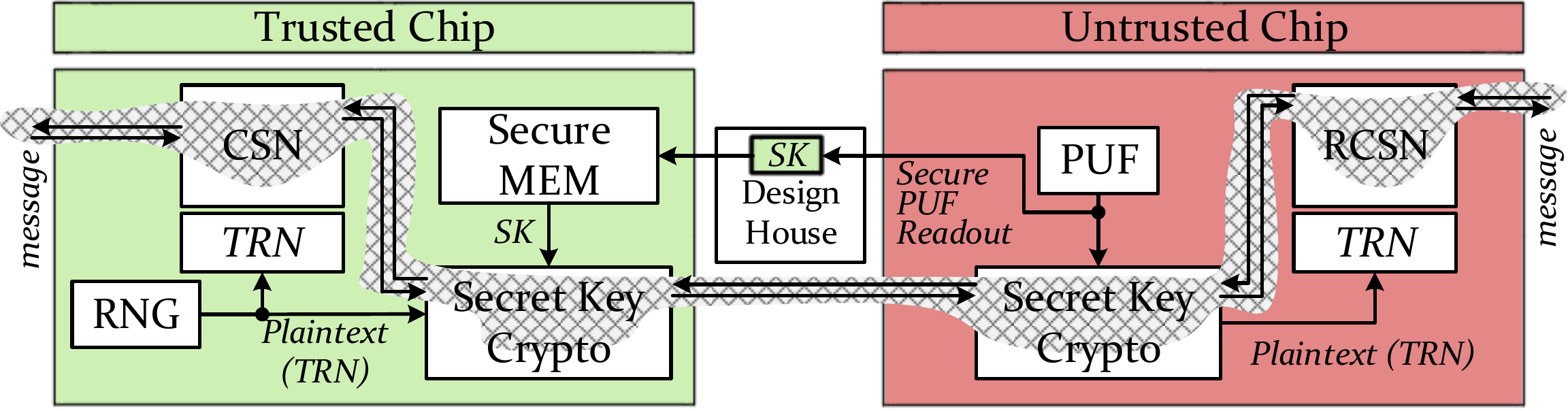}
    \caption{Double Cipher mode of Cryptography (DCC)}
  \end{subfigure}
    \begin{subfigure}[b]{\columnwidth}
    \includegraphics[width=\textwidth]{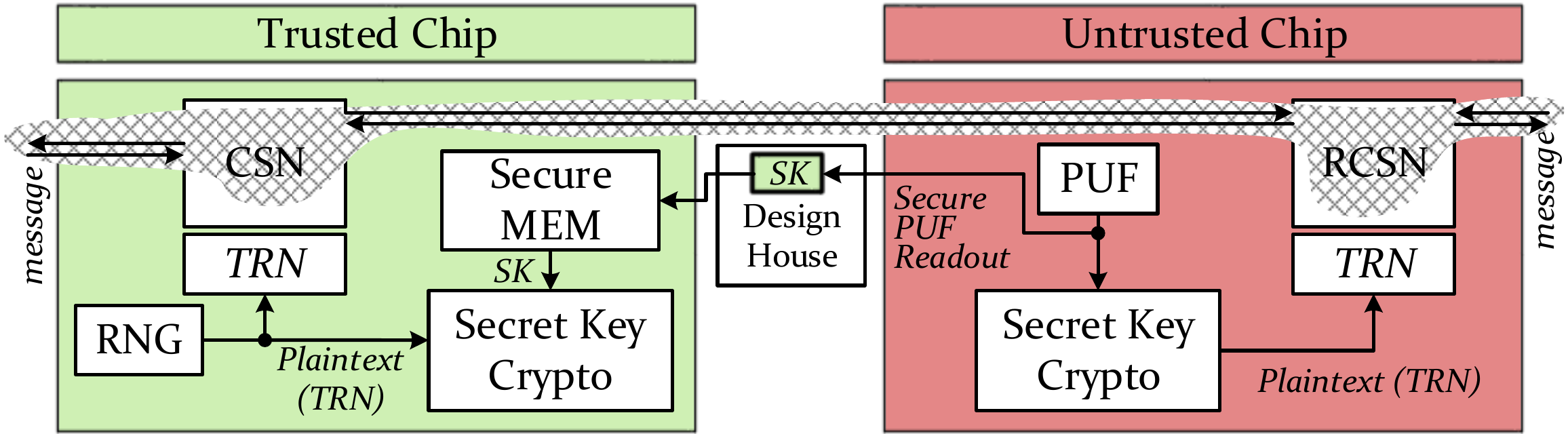}
    \caption{Leaky Cipher mode of Cryptography (LCC)}
  \end{subfigure}
   
    \caption{Modes of Encrypted Communication in COMA \vspace{-3mm}}
    %\vspace{-5mm}
    \label{DCC_LCC}
\end{figure}

In 2.5D-COMA, the untrusted chip(s) is used as an accelerator, and for safety reasons should not be able to directly communicate to the outside world. Hence, all communication to/from the untrusted chip must go through the trusted chip. In addition, it is possible that the computation, depending on the sensitivity of processed data, is divided between the trusted and untrusted chips. Hence, there is a need for constant communication between the trusted and untrusted chips. The communication needed is sometimes for limited but highly sensitive data, and sometimes for vast amounts of less sensitive data. As illustrated in Fig. \ref{DCC_LCC}, the proposed architecture is designed to provide two hybrid means of encrypted communication : (1) Double-Cipher Communication (DCC) as ultra-secure communication, and (2) Leaky-Cipher Communication (LCC) as ultra-fast communication mechanism.

%\vspace{-2mm}
\subsubsection{\textbf{Double-Cipher Communication (DCC)}}
\vspace{-2mm}
As shown in Fig. \ref{DCC_LCC}(a), \textbf{DCC} is implemented by passing each message through both CSN-RCSN and the secret key cryptography engine, where the TRN used in CSN-RCSN is renewed every $U$ cycles. DCC provides the ultimate protection against side-channel attacks. In DCC mode, two necessary requirements for mounting a side channel attack are eliminated. The side channel attack aims to break the cryptography system by analyzing the leaked side channel information for different \emph{input patterns}. Hence, (1) the degree of correlation between the input and the leaked side-channel information, and (2) the intensity of side-channel variation, are important. In COMA, the attacker cannot control the input to the secret-key cryptography. In addition, the input to the CSN is randomized using a TRN and then passed to the secret-key cryptography, removing the correlation between leaked side channel info (from secret-key cryptography) and the original input to the CSN. Additionally, the secret-key cryptography engine is side-channel protected to pass a t-test \cite{gilbert2011testing}. So, the intensity and variation in side-channel information is significantly reduced, making the DCC an extremely difficult attack target. 

% %%\vspace{5pt}
% \begin{figure}
% \centering
% % \subfloat[]{{\includegraphics[width=0.8\columnwidth]{ExTruDYN1.pdf} }} \newline %\vspace{-5pt}
% \includegraphics[width=\columnwidth]{images/leaky_trans.pdf} 
% %\vspace{-18pt} 
% \caption{The Impact of TRN Update on LCC mode. %\vspace{-14pt}}
% \label{expdyn}
% %\vspace{-2mm}
% \end{figure}

\begin{figure*}[t]
\centering
\includegraphics[width = 2.1\columnwidth]{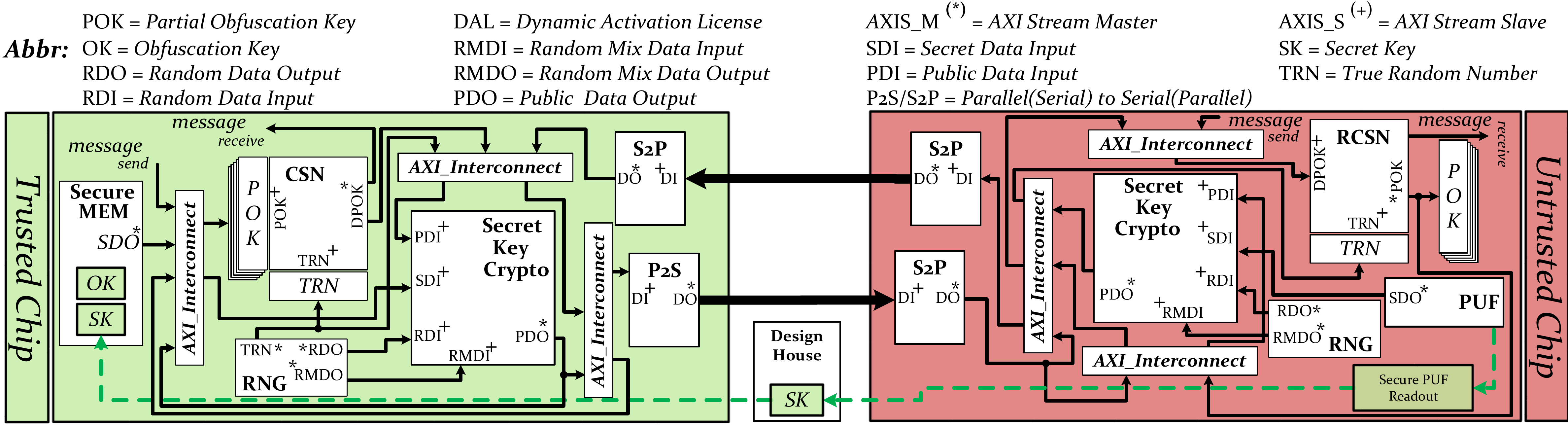}
% %\vspace{-3mm}
\caption{\emph{2.5D-COMA} Architecture. \vspace{-3mm}}
\label{3d_detail_arch}
%\vspace{-4mm}
\end{figure*}

%\vspace{-2mm} 
\subsubsection{\textbf{Leaky-Cipher Communication} (\textbf{LCC})}
\vspace{-2mm}

\textbf{LCC} is a fast and energy efficient mode of communication between the trusted (or remote device) and the untrusted chip. As illustrated in Fig. \ref{DCC_LCC}(b), in this protocol, the CSN-RCSN pair is used for exchanging data. The secret key cryptography engine is used to transmit a TRN from one chip to the other. Since the throughput of TRNG is the bottleneck point compared to the performance of CSN-RCSN, the TRNG is used as a seed generator to the PRNG (which offers higher performance) on both sides, Hence, in LCC mode, PRNG is used to configure the CSN-RCSN to avoid any performance degradation on transmitting data. For $U$ consecutive cycles, the PRNG is kept idle allowing the CSN to use the same PRNG output for $U$ cycles. It not only reduces the power consumption of PRNG and TRNG, it also provides faster communication in LCC mode. However, using this model of communication is prone to algebraic and SAT attacks as each communicated message leaks some information about the TRN used to configure the CSN-RCSN pair. If an attacker can control the message and observe the output of the CSN, each communicated message leaks some information about the key, reducing its security. Extracting the key from such observations is possible by various attack models, including Satisfiability attacks. Hence, an attacker with enough time and enough traces could extract the TRN and retrieve the communicated messages. Preventing such attacks poses a minimum limit to $U$ (the update frequency of the PRNG). $U$ should be small to prevent SAT and other trace-based learning or analysis attacks, but large enough to be energy efficient. In Section~\ref{results}, we deploy a SAT attack against LCC and will further elaborate on the required TRN update frequency.

% As suggested previously, the untrusted chip is activated in DCC mode. The major steps for unlocking the obfuscated circuit in the untrusted chip using DCC communication flow is as follow: \textcolor{red}{Fig. \ref{3d_detail_arch} ??}

% \begin{algorithm}[h]
% \footnotesize
% % \caption{Activating the obfuscated circuit in the untrusted chip }\label{remote_activation}
% \begin{algorithmic}[b]
% \State T $\gets$ Trusted chip or Activation Manager
% \State U $\gets$ Untrusted chip 
% \State OK$_{size}$ $\gets$ T reads Obfuscation Key (OK) size from memory
% \State T Sends a TRN request to the U
% \State U generates the TRN and loads it into its RCSN
% \State U encrypts and sends the TRN to T
% \State T decrypts the cipher and loads TRN into CSN
% \State Set s = CSN$_{size}$
% \State Set Size$_{sent}$ = 0
% \While {(Size$_{sent}$ $\leq$ OK$_{size}$) }
%   \State T reads $s$ bits of OK from memory \Comment{Partial OK (POK)} 
%   \State T encrypts (in DCC mode) and sends the packet to U
%   \State U Decrypts the packet and restore the partial OK 
%   \State U loads the partial OK into activation scan chain registers
%   \State Size$_{sent}$ $\gets$ Size$_{sent}$ + CSN$_{size}$ %Size$_{message}$
% \EndWhile
% \end{algorithmic}
% \end{algorithm}

\vspace{-2mm} 
\subsection{\textbf{R-COMA:} for Protecting IoT devices}
\vspace{-2pt} 

The R-COMA architecture in the untrusted chip is identical to that of 2.5D-COMA. However, the trusted chip is replaced with a remote key management service. The R-COMA provides a mechanism for an IP owner to remotely activate parts or entire functionality of the hardware. Similar to 2.5D-COMA, the DAL is different from chip to chip and from activation to activation. In R-COMA, the obfuscation unlock key is stored in a central database, while the CSN, the TRNG for configuring CSN-RCSN, and the secret key cryptography engine are implemented in software. 

In \emph{R-COMA}, an authentication server (AS) first securely receives the PUF-based SK from the untrusted chip. Then, it generates a TRN and sends it to the untrusted chip for RCSN configuration. Then, the AS starts sending the obfuscation key (OK). For the activation phase, the communication is double encrypted and authenticated using the CSN-RCSN and side-channel protected cryptography engine. Each COMA-protected device needs to be registered with the AS to receive the obfuscation key. The registration is done using the PUF-ID of the untrusted chip. Hence, the PUF is used for both authentication and generation of the secret key for communication. In R-COMA, the generation of \emph{DAL} is granted after PUF authentication, and is based on the generated TRN, and the stored OK, which is generated at design time. The generation of DAL is algorithmic and takes linear time.

%\vspace{-2pt} 
\subsection{Implementation Detail of COMA} \label{Macros}
\vspace{-2pt} 

Fig. \ref{3d_detail_arch} captures the overall architecture of COMA and relation and connectivity of its macros. As discussed, COMA supports both key-management and secure data communication. Based on the selected mode of communication (LCC/DCC), the message passes through \{CSN {\scriptsize{$\rightarrow$}} RCSN\} \emph{or} \{CSN {\scriptsize{$\rightarrow$}} encryption {\scriptsize{$\rightarrow$}} decryption {\scriptsize{$\rightarrow$}} RCSN\}. RNG, which contains both TRNG and PRNG, is used in both sides. In the trusted chip, RNG is used for implementing side-channel protected cryptography engine, as well as generating the configuration of the CSN-RCSN (TRN). In the untrusted side, it is used only for implementing the side-channel protected cryptography engine. Finally, PUF is engaged in the untrusted chip for both unique IC authentication and for generation of the secret key for encryption. As shown in Fig. \ref{3d_detail_arch}, all modules employ an AXI-stream interface to maximize the simplicity of the overall design, and minimize the overhead incurred by the controller of the top module in each side. The description of the behavior of each macro in COMA is provided next:

\begin{figure}[t]
\centering
\includegraphics[width = \columnwidth]{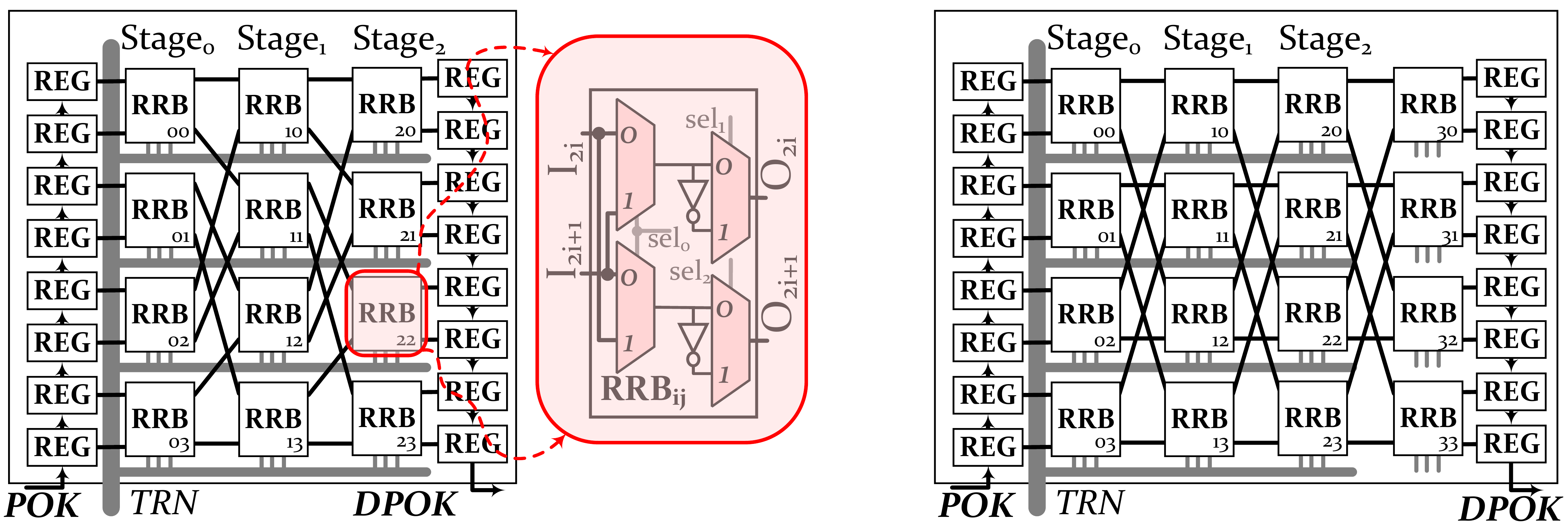}
(a) \hspace{130pt} (b)
\vspace{-2mm}
\caption{Logarithmic Network (a) Omega-based Blocking, (b) $LOG_{8, 1, 1}$ near Non-blocking. \vspace{-3mm}}
\label{csn_arch}
%\vspace{-5mm}
\end{figure}

%\vspace{-7pt} 
\subsubsection{\textbf{Configurable Switching Network (CSN)}} \label{CSN_section}
\vspace{-2pt} 

The CSN is a logarithmic routing network that could route the signals at its input pins to its output pins while permuting their order and possibly inverting their logic levels based on its configuration. Fig. \ref{csn_arch}(a) captures a simple implementation of an 8-by-8 CSN using \emph{OMEGA} \cite{ahmadi1989survey} network. The network is constructed using routing elements, denoted as Re-Routing Blocks (RRB). Each RRB is able to possibly invert and route each of the input signals to each of its outputs. The number of RRBs needed to implement this simple CSN for $N$ inputs ($N$ is a power of 2) is simply $N/2*logN$. Each CSN should be paired with an RCSN. The RCSN, is simply constructed by flipping the input/output pins of RRB, and treating the CSN input pins as its output pins and vice versa.

The \emph{OMEGA} network along with many other networks of such nature (Butterfly, etc.) are blocking networks \cite{ahmadi1989survey}, in which we cannot produce all permutations of input at the network's output pins. This limitation significantly reduces the ability of a CSN to randomize its input. Also, we will show that a blocking CSN can be easily broken by a SAT attack within few iterations. 

Being a blocking or a non-blocking CSN depends on the number of stages in CSN. Since no two paths in an RRB are allowed to use the same link to form a connection, for a specific number of RRB columns, only a limited number of permutations is feasible. However, adding extra stages could transform a blocking CSN into a strictly non-blocking CSN. Using a strictly non-blocking CSN not only improves the randomization of propagated messages through the CSN, but also improves the resiliency of these networks against possible SAT attacks for extraction of a TRN used as the key for a CSN-RCSN cipher. A non-blocking logarithmic network could be represented using $LOG_{n, m, p}$, where $n$ is the number of inlets/outlets, $m$ is the number of extra stages, and $p$ indicates the number of copies \emph{vertically cascaded} \cite{shyy1991log}. 

According to \cite{shyy1991log}, to have a strictly non-blocking CSN for an arbitrary $n$, the smallest feasible values of $p$ and $m$ impose very large area/power overhead. For instance, for $n=64$, the smallest feasible values, which make it strictly non-blocking, are $m=3$ and $p=6$, which means there exists more than $5\times$ as much overhead compared to a blocking CSN with the same $n$, resulting in a significant increase in the area and delay overhead. To avoid such large overhead, we employ a \emph{close to non-blocking CSN} described in \cite{shyy1991log} to implement the CSN-RCSN pair. This network is able to generate not all, but \emph{almost all} permutations, while it could be implemented using a $LOG_{n, log_2(n) - 2, 1}$ configuration, meaning it needs $log_2(n) - 2$ extra stages and no additional copy. Fig. \ref{csn_arch}(b), demonstrates an example of such a close-to-non-blocking CSN with $n = 8$. In the results section, we demonstrate that using these close-to-non-blocking CSNs enhances the resiliency of a CSN against SAT attack, even in small sizes of CSNs with significantly lower power, performance and area (PPA) overhead. 

%\vspace{-11pt} 
\subsubsection{\textbf{Authenticated Encryption with Associated Data}} \label{AEAD}
\vspace{-2pt} 

The Authenticated Encryption with Associated Data (AEAD) is used in the DCC mode for communicating messages, and in the LCC mode for the initial transmission 
% a sporadic update 
of the CSN-RCSN key (TRN). Authenticated ciphers incorporate the functionality of confidentiality, integrity, and authentication. The input of an authenticated cipher includes Message, Associated Data (AD), Public Message Number (NPUB), and a secret key. The ciphertext is generated as a function of these inputs. A Tag, which depends on all inputs, is generated after message encryption to assure the integrity and authenticity of the transaction. This tag is then verified after the decryption process. The choice of AEAD could significantly affect the area overhead of the solution, the speed of encrypted communication, and the extra power consumption. To show the performance, power, and area trade-offs, we employ two AEAD solutions: a NIST compliant solution (AES-GCM), and a promising lightweight solution (ACORN).

AES-GCM is the current National Institute of Standards and Technology (NIST) standard for authenticated encryption and decryption as defined in \cite{dworkin2007recommendation}. ACORN is one of two finalists of the Competition for Authenticated Encryption: Security, Applicability, and Robustness (CAESAR), in the category of lightweight authenticated ciphers, as defined in \cite{wu2016acorn}. An 8-bit side-channel protected version of AES-GCM and a 1-bit side-channel protected version of ACORN are implemented as described in \cite{diehl2018face}. Both implementations comply with lightweight version of the CAESAR HW API \cite{homsirikamol2015gmu}.

Our methodology for side channel resistant is threshold implementation (TI), which has wide acceptance as a provably secure Differential Power Analysis (DPA) countermeasure~\cite{nikova2006threshold}. In TI, sensitive data is separated into shares and the computations are performed on these shares independently. TI must satisfy three properties: 1) Non-completeness: Each share must lack at least one piece of sensitive data, 2) Correctness: The final recombination of the result must be correct, and 3) Uniformity: An output distribution should match the input distribution. To ensure uniformity, we refresh TI shares after non-linear transformations using randomness. We use a hybrid 2-share/3-share approach, where all linear transformations in each cipher are protected using two shares, which are expanded to three shares only for non-linear transformations. 

To verify the resistance against DPA, we employ the Test Vector Leakage Assessment  methodology in \cite{gilbert2011testing}. We leverage a "fixed versus random" non-specific t-test, in which we randomly interleave first fixed test vectors and then randomly-generated test vectors, leading to two sequences with the same length but different values. Using means and variances of power consumption for our fixed and random sequences, we compute a figure of merit $t$. If $|t| > 4.5$, we reason that we can distinguish between the two populations and that our design is leaking information. 
The protected AES-GCM design has a 5-stage pipeline and encrypts one 128-bit input block in 205 cycles. This requires 40 bits of randomness per cycle. In ACORN-1, there are ten 1-bit TI-protected AND-gate modules, which consume a total of 20 random reshare, and 10 random refresh bits per state update. In a two-cycle architecture, 15 random bits are required per clock cycle.

%\vspace{-3mm} 
\subsubsection{\textbf{Random Number Generator (RNG)}} \label{RNG}
\vspace{-2pt} 

An RNG unit is required on both sides to generate random bits for side channel protection of AEAD units, a random public message number (NPUB) for AEAD, and TRNs for CSN-RCSN. We adopted the ERO TRNG core described in \cite{petura2016survey}, which is capable of generating only 1-bit of random data per over 20,000 clock cycles. In our TI implementations, AES-GCM needs 40 and ACORN 15 bits of random data per cycle. So, we employed a hybrid RNG unit combining the ERO TRNG with a Pseudo Random Number Generator (PRNG). TRNG output is used as a 128-bit seed to PRNG. The PRNG generates random numbers needed by other components. The reseeding is performed only once per activation.

The choice of PRNG depends on the expected performance and overhead. To support COMA, we adopted two different implementations of PRNG: (1) AES-CTR PRNG, which is based on AES, is compliant with the NIST standard SP 800-90A, and generates 12.8 bits per cycle. (2) Trivium based PRNG, which is based on the Trivium stream cipher described in \cite{de2005trivium}. The Trivium-based PRNG is significantly smaller in terms of area and much faster than AES-CTR PRNG. It can generate 64 bits of random data per cycle, however, it is not compliant with the NIST standard.

%\vspace{-2mm} 
\subsubsection{\textbf{PUF and Secure PUF Readout}} \label{secure_puf_readout}
\vspace{-2pt} 

The response of the PUF to a challenge selected randomly by Enrollment Authority (SoC designer) is used as the secret key in AEAD. Hence, the readout of the PUF-response should be protected. The simplest solution for the safe readout of a PUF-generated key is to enable the readout by burning one set of fuses, and disabling it by burning a second set of fuses.
%However, this solution requires the PUF readout to take place in a trusted foundry. 
However, this solution, especially when combined with a weak PUF, is not likely to be resistant against the untrusted foundry, which may possibly burn the first set of fuses, read out PUF key, and then repair fuses before releasing the chip. To avoid this problem, we implement a lightweight one-sided public key cryptography (encryption only) based on Elliptic-Curve Cryptography (ECC). Considering the PUF readout is a one-time event, the  performance of the public-key cryptography engine is not critical.

In order to prevent any attempts at fully characterizing a PUF in the untrusted foundry, only strong PUFs, e.g. an arbiter PUF, are considered. The secure readout of the PUF key is allowed only at the device enrollment time, in the secure facility. During the secure readout, the strong PUF is fed with multiple challenges selected by the Enrollment Authority. The corresponding PUF responses are encrypted by the untrusted chip using the public key of the Enrollment Authority, that is embedded in the chip layout or stored in the one-time programmable memory. Only the Enrollment Authority has access to the decrypted responses. Afterwords, one of the previously applied challenges is randomly selected and used for the generation of the secret key. This challenge is then hardwired on the untrusted chip, and the PUF response to that challenge is recorded by the Enrollment Authority. This PUF response is then stored in the secure memory of the trusted chip in 2.5D-COMA, or in the secure cloud directory in R-COMA. This process makes each PUF key unique to a given device, and resistant against any unauthorized readout by the untrusted foundry.

Still, additional precautions must be taken to protect this scheme against an attack aimed at replacing a real PUF by a pseudo-PUF, generating randomly looking responses that can be easily calculated by an attacker. An example of such a pseudo-PUF may be a lightweight symmetric-key cipher, with a fixed key known to the untrusted party, encrypting each challenge and outputing a ciphertext as the PUF response.

Such pseudo-PUF should be treated as a Trojan and detected by Enrollment Authority using the best known anti-Trojan techniques, e.g., those based on the measurement and analysis of the power consumption during the operation of the device \cite{agrawal2007trojan}. Additional methods may be used to differentiate the outputs of a strong PUF from encrypted data, e.g., using known correlations between the PUF responses corresponding to closely-related challenges, such as challenges differing on only one bit position, or being mutual complements of each other \cite{delvaux2014attacking}. These kinds of PUF-health tests may be specific to a particular strong PUF type, e.g., to an arbiter PUF, and will be the subject of our future work.

%\vspace{-2mm} 
\section{COMA Resistance against various Attacks}  \label{attacks}
\vspace{-2mm} 
\subsection{Assumed Attacker Capabilities}
\vspace{-2mm} 
Different sources of vulnerability are considered in this section to demonstrate the COMA security. The attacker can be an adversary in the manufacturing supply chain, and has access to either the reverse engineered or design house-generated netlist of the COMA-protected untrusted chip. The attacker can purchase an activated COMA-protected IC from the market. The attacker can monitor the side channel information of chips at or post activation. The attacker can observe the communication between untrusted and trusted (or remote manager) chips and could also alter the communicated data. An Attack objective may be (1) extracting the obfuscation key (OK), (2) illegal activation of the obfuscated circuit without extracting the key, (3) extracting the long-term secret key (SK), (4) extracting short-term CSN keys (TRNs), (5) eavesdropping on messages exchanged between the untrusted chip and the external sources, (6) removing the COMA protection, or (7) COMA-protected IC overproduction.  

\begin{figure*}[t]
  \centering  
  \begin{subfigure}[b]{0.25\textwidth}
    \includegraphics[width=0.9\textwidth]{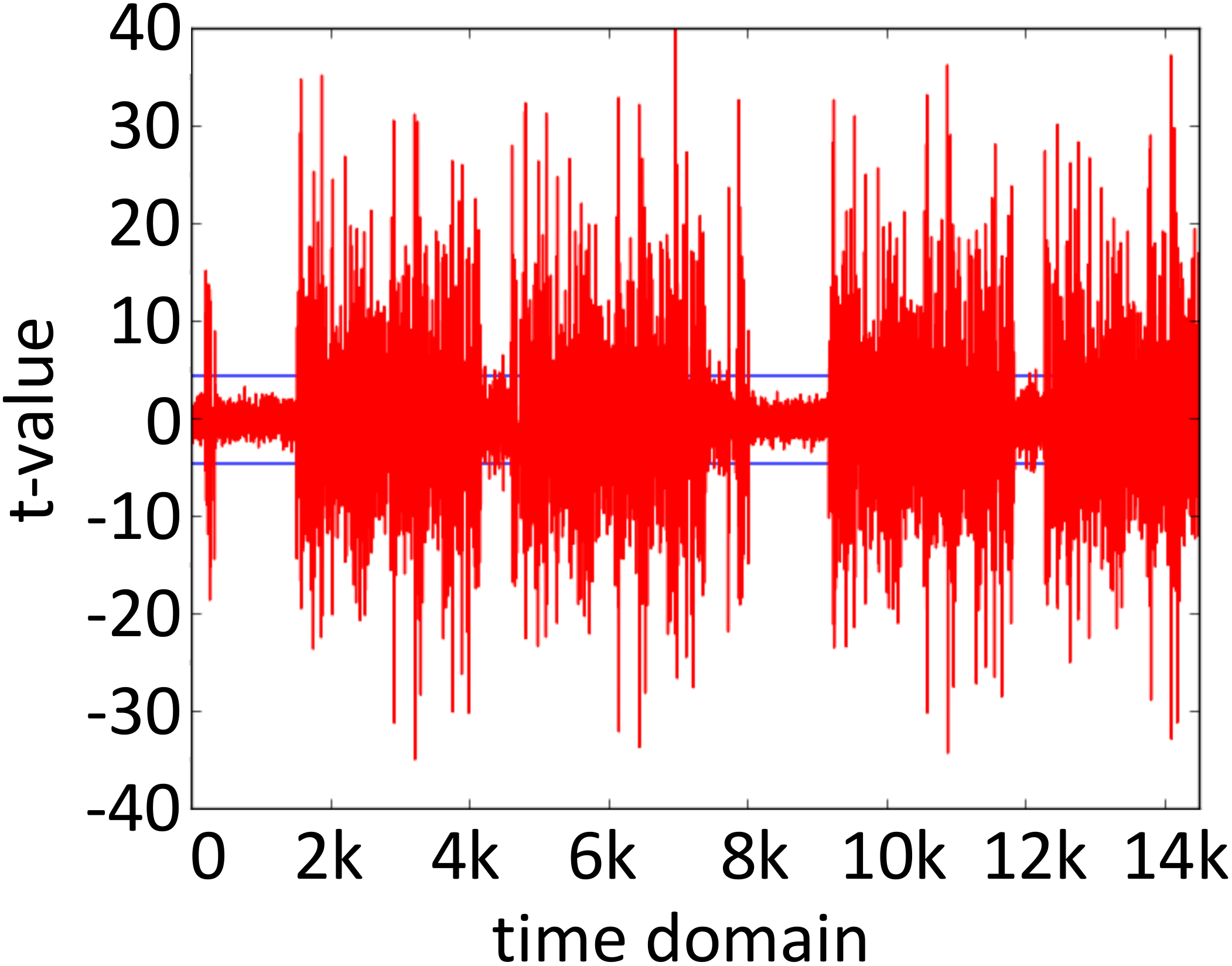}
    \caption{AES-GCM UnPr}
    \label{fig:aesgcmunpr}
  \end{subfigure}
  %\hspace{1em}
  %\hfill
  \begin{subfigure}[b]{0.25\textwidth}
    \includegraphics[width=\textwidth]{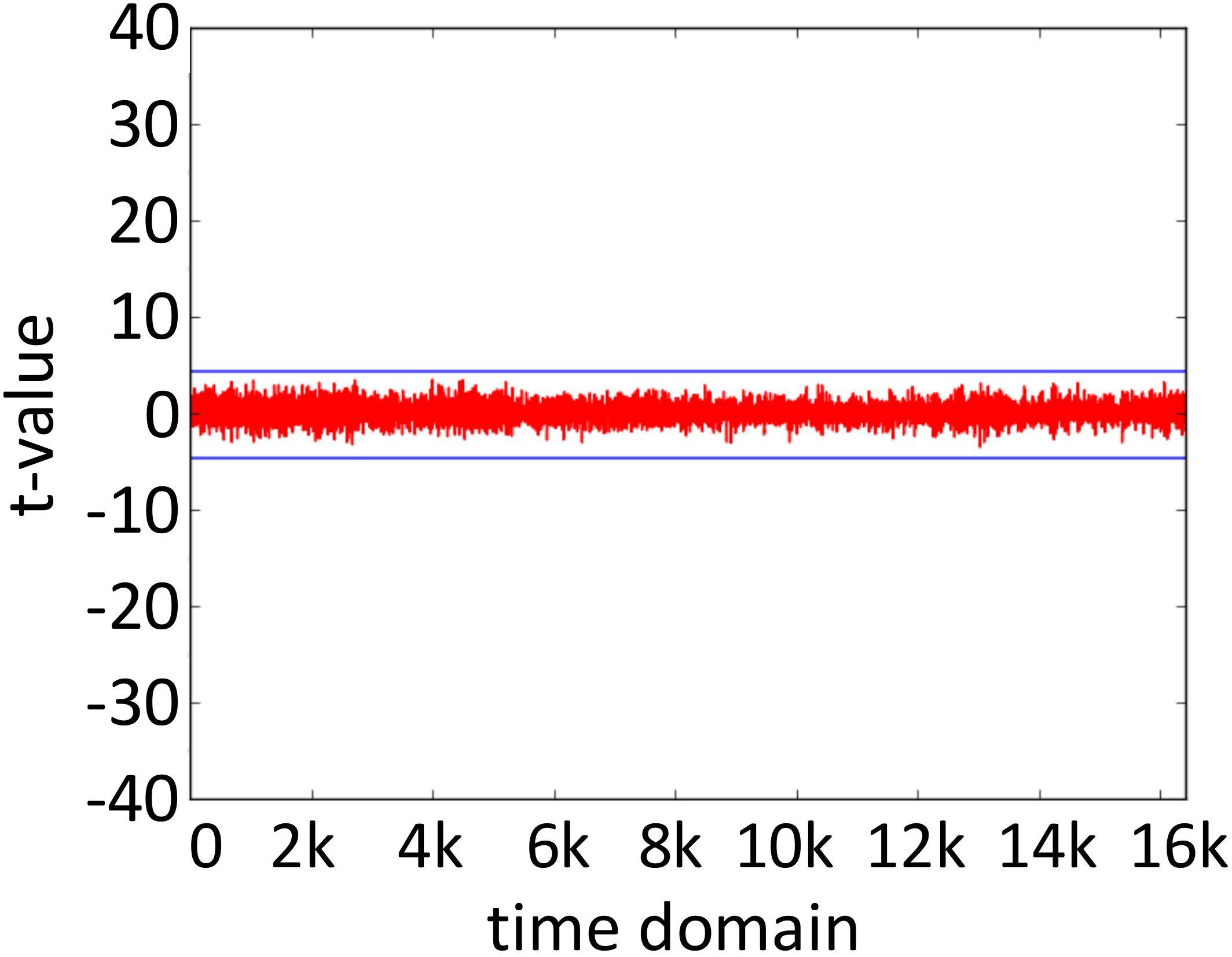}
    \caption{AES-GCM Pr}
    \label{fig:aesgcmpr}
  \end{subfigure}
  \begin{subfigure}[b]{0.24\textwidth}
    \includegraphics[width=\textwidth]{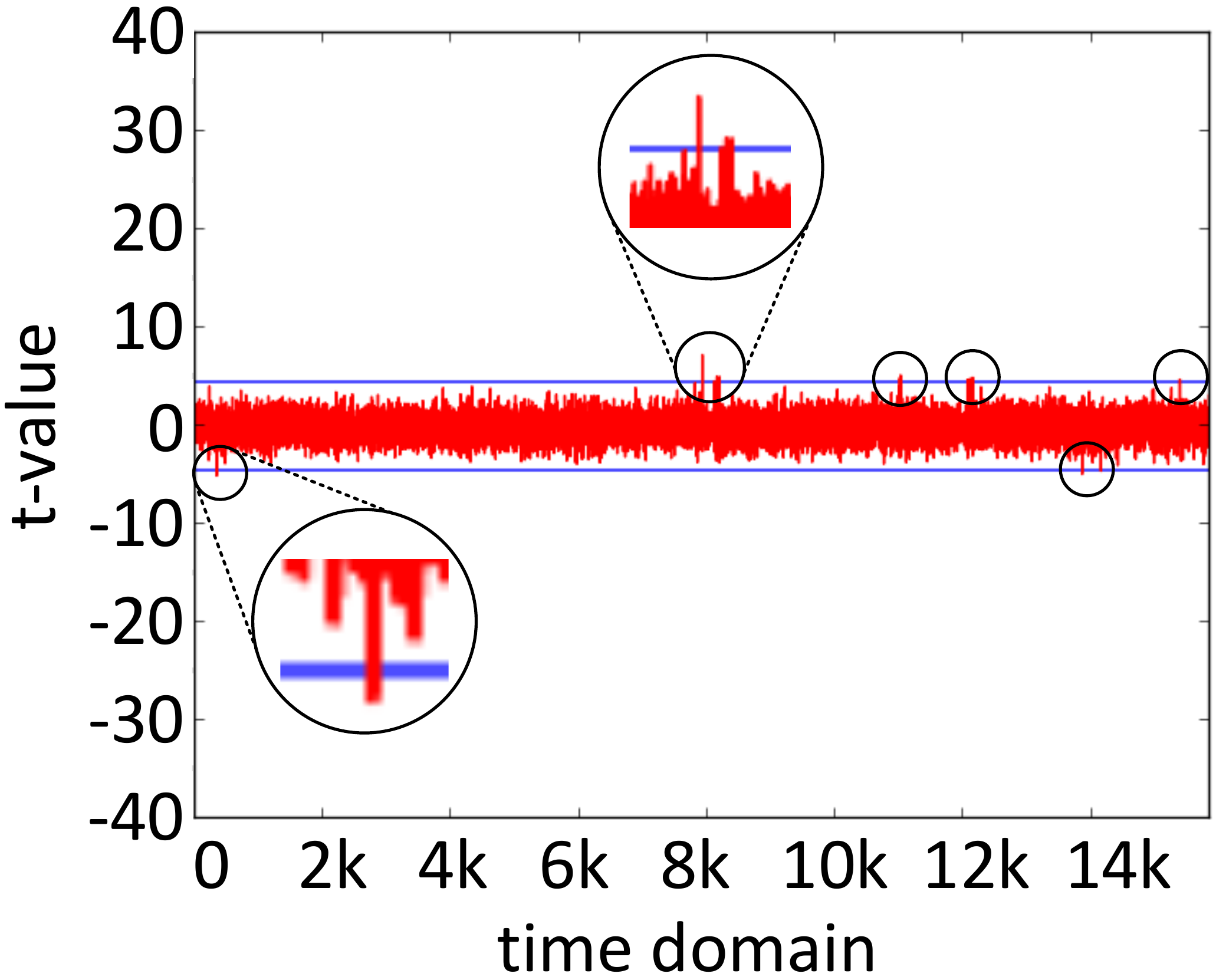}
    \caption{ACORN UnPr}
    \label{fig:acornunpr}
  \end{subfigure}
  \begin{subfigure}[b]{0.24\textwidth}
    \includegraphics[width=\textwidth]{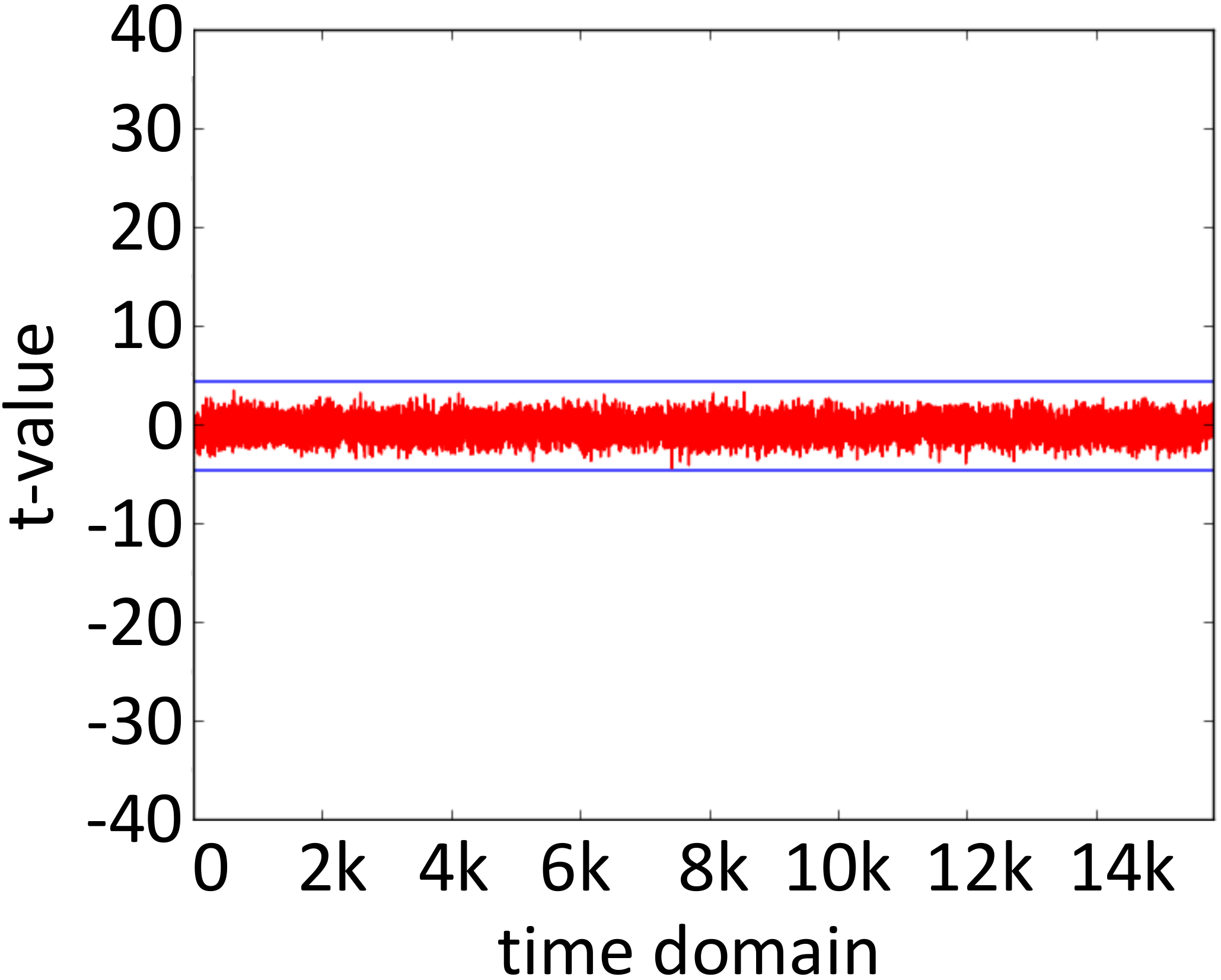}
    \caption{ACORN Pr}
    \label{fig:acornpr}
  \end{subfigure}
    \vspace{-4pt}
    \caption{The t-test results for unprotected (UnPr) and protected (Pr) implementations of AES-GCM and ACORN.\vspace{-4mm}}
    % %\vspace{-15pt}
  \label{fig:ttest}
\end{figure*}

%\vspace{-4mm} 
\subsection{\textbf{Side Channel Attack (SCA)}}
\vspace{-2pt} 

The objective of SCA on COMA is to extract either the secret key (SK) used by AEAD or the TRN used by CSN. Extracting a SK is sufficient to break the obfuscation; extracting a TRN reveals only messages sent in the LCC mode.

DCC significantly increases the SCA difficulty, since (1)\,the AEAD is side-channel protected, and (2) the attacker loses access to the input of AEAD. Fig. \ref{fig:ttest} captures our assessment of side channel resistance of AEAD using a t-test for unprotected and protected implementations of AES-GCM and ACORN  \cite{diehl2018comparison}. As illustrated, both implementations pass the t-test, indicating increased resistance against SCA. On the other hand, the inability to control the input to AEAD comes from the COMA requirement of encryption in the DCC mode where a message first passes through the CSN. Hence, there exists no relation between the power consumption of the AEAD and the original input due to CSN randomization. CSN power consumption is also randomized as it is a function of $n$ inputs (possibly known to the attacker) and $3n \times (log_2 n -1)$ TRN inputs unknown to the attacker, while the TRN is repeatedly updated based on the value of $U$. Note that during the physical design of COMA, the side channel information on power and voltage noise (IR drop) could be further mitigated using timing aware IR analysis \cite{vakil2019ir}, and voltage noise aware clock distribution techniques \cite{7987531, 8009179}. 

The LCC mode is prone to side-channel, algebraic, and SAT attacks aimed at extracting the TRN. However, the attack must be carried out in a limited time while the TRN of the CSN/RCSN is unchanged. As soon as the TRN is renewed, the previous side-channel traces or SAT iterations are useless. The period of TRN updates ($U$) introduces a trade-off between energy and security and can be pushed to maximum security by changing the TRN for every new input. In section \ref{TRN_Perf_LCC} we investigate the time required to break the LCC using side-channel or SAT attack and accordingly define a safe range for $U$ to prevent such attacks. 

%\vspace{-4mm} 
\subsection{\textbf{Reverse Engineering}} 
\vspace{-2pt}

% COMA is not an obfuscation solution but an obfuscation key-management scheme. Hence, in COMA, we assume that the netlist is obfuscated with a SAT-Hard obfuscation scheme, such as SARLock, Anti-SAT, or SFLL, to name a few \cite{yasin2016sarlock, xie2016mitigating, yasin2017provably}. The security objectives of COMA is to keep both the obfuscation key and communication data secure against invasive and non-invasive attacks. 
In COMA, reverse engineering (RE) to extract the secret key from layout is useless as the secret key is not hardwired in the design and is generated based on PUF. RE to extract the key from memory in an untrusted chip is no longer an option as the key is not stored in the untrusted chip. RE to extract the key from the trusted chip's memory is limited by the difficulty of tampering with secure memory in the trusted technology.

%\vspace{-4mm}
\subsection{\textbf{Algebraic Attacks}} 
\vspace{-4pt}

Algebraic attacks involve (a) expressing the cipher operations as a system of equations, (b) substituting in known data for some variables, and (c) solving for the key. AES-GCM and ACORN have been demonstrated to be resistant against all known types of algebraic attacks, including linear cryptanalysis. Therefore, in the absence of any new attacks, the DCC mode is resistant against algebraic attacks. Using CSN and RCSN for fast encryption is new and requires more analysis. CSN can be expressed as an affine function of the data input $x$, of the form $y=A\cdot x + b$, where $A$ is an $n \times n$ matrix and $b$ is an $n \times 1$ vector, with all elements dependent on the input TRN. Although recovering $A$ and $b$ is not equivalent to finding the TRN, it may enable the successful decryption of all blocks encrypted using a given TRN. We protect against this threat in two ways: (1) The number of blocks encrypted using a given TRN is set to the value smaller than $n$, which prevents generating and solving a system of linear equations with $A$ and $b$ treated as unknowns, (2) We partially modify the TRN input of CSN with each block encryption (by a simply shifting the input TRN bits), so the values of $A$ and $b$ are not the same in any two encryptions, without the need of feeding CSN with two completely different TRN values.  

%\vspace{-2mm} 
\subsection{\textbf{Counterfeiting and Overproduction}} 
\vspace{-3pt}

COMA can be used to prevent the resale of used ICs, usage of illegal copies, and reproduction of a design. During packaging and testing, each COMA protected IC is first tested and then is matched with a trusted chip. So, the untrusted chip can only be activated by the matched trusted chip or the registered remote manager. Building illegal copies that work without the secure chip (or remote activation) and reproduction of the design requires successful RE. Blind reproduction is useless as its activation requires a matching trusted chip or passing PUF authentication of a remote manager. By receiving one or more DALs for testing, the manufacturer cannot activate additional IPs as the DAL changes from activation to activation. 

%\vspace{-3mm} 
\subsection{\textbf{Removal attacks}} 
\vspace{-3pt}

Removal of the TRNG fixates the DAL and breaks the LCC mode. In DCC mode, it gives an attacker control over the input to the AEAD, increasing the chances of SCA on the cryptography engine. NIST standard SP 800-90B \cite{barker2012recommendation} dictates that continuous health testing must be performed on the TRNG. These tests include repetition counting to detect catastrophic failure and adaptive proportion testing to detect loss of entropy. Removal of the TRNG would be detected as this would result in insufficient entropy to satisfy the health test, assuming the test is implemented on the trusted chip. Removal of COMA architectural modules makes the chip non-functional as COMA is not a wrapper architecture, but a fused one. Complete removal of COMA requires successful RE. Removing the PUF can be made challenging by using a strong PUF, with a large number of challenge-response pairs. Replacing such a PUF with a deterministic function is challenging as such functions are likely to have a substantially different area and power, making them detectable.

\begin{table}[htb!]
\scriptsize
\centering
\caption{Main features of the two proposed COMA variants. \vspace{-2mm}}
\label{COMA_configss}
\setlength\tabcolsep{5pt} % default value: 6pt
\scalebox{0.9}{

\begin{tabular}{@{} l  *3c @{}}
\toprule
\multicolumn{1}{c}{Feature} & COMA1 & COMA2 \\
\midrule
AEAD    & AES-GCM & ACORN \\
PRNG    & AES-CTR & Trivium \\
BUS Width & 8 & 8 \\
Pins used for Communication & 8 & 8 \\
CSN-RCSN Size & 64 & 64 \\
%PUF & Memory PUF & Memory PUF \\ 
Trusted Memory & 4 Kbits & 4 Kbits \\
C$_{fix}$: initialization overhead (cycles) & 10,492 & 20,452 \\
C$_{byte}$: cycles needed for encrypting each byte & 72 & 17  \\ 
PRNG$_{perf}$: Throughput of generating PRN & $128bit / 10cycles$ & $64bit / cycle$  \\ \bottomrule
\end{tabular}
}
%%\vspace{-2mm}
\end{table}

%\vspace{-3mm}
\section{COMA Implementation Results} \label{results}
\vspace{-2mm}

For evaluation, all designs have been implemented in VHDL and synthesized for both FPGA and ASIC. For ASIC implementation we used Synopsys generic 32nm educational libraries. For FPGA verification, we targeted a small FPGA board, Digilent Nexys-4 DDR with Xilinx Artix-7 (XC7A100T-1CSG324). 

%\vspace{-3mm}
\subsection{COMA Area Overhead} \label{COMA_overhead}
\vspace{-2mm}

\begin{table}
\scriptsize 
\centering
\caption{Resource Utilization of the COMA Architecture for NIST-compliant and lightweight solution. \vspace{-2mm}}
\label{FPGA_overhead}
\setlength\tabcolsep{9.2pt} % default value: 6pt
% \begin{tabular}{|c|c|c|c|c|c|c|c|}
\scalebox{0.87}{

\begin{tabular}{@{} l *7c @{}}
\toprule 
         {Name} & \multicolumn{3}{c}{AES-GCM+AES-CTR} & &\multicolumn{3}{c}{ACORN+Trivium} \\ 
        \cmidrule(lr){2-4}
        \cmidrule(lr){6-8}
                & Slice & LUT  & FF   &  &Slice & LUT & FF \\ \midrule
                \multicolumn{8}{c}{TRUSTED}\\ \midrule
AEAD\_EXT       & 1,336   & 3,804  & 4,432  &  &333 & 1,067 & 591 \\ 
RNG           & 712   & 2,226   & 618  &  &215 & 601 & 450 \\
CSN             & 257   & 540   & 739  &  & 257 & 540  & 739 \\
Others          & 149     & 345      & 144  &   & 149 & 345 & 144  \\ \midrule
                \multicolumn{8}{c}{UNTRUSTED}\\ \midrule
AEAD\_EXT       & 1,336   & 3,804  & 4,432  &  &333 & 1,067 & 591 \\ 
RNG             & 738   & 2,352   & 628  &  &241 & 683 & 460 \\
RCSN            & 252   & 607   & 737  &  &252 & 607  & 737 \\
ECC & 563 & 1569 & 1161 & & 563 & 1569 & 1161  \\
PUF \cite{machida2015implementation} & 177 & --- & --- & & 177 & --- & ---  \\
Others          &209       & 359      & 257     &  &209    & 359     & 257  \\ \bottomrule
\end{tabular}
}
\begin{flushleft}
\vspace{-6pt}
\textit{On Xilinx Artix-7 (XC7A100T-1CSG324) FPGA. }
\end{flushleft}
%\vspace{-10pt}
\end{table}
% %\vspace{-15pt}
% %\vspace{-4mm}

We implemented two variants of COMA architecture: a NIST compliant solution (denoted by COMA1) and a lightweight solution (denoted by COMA2). The AEAD and PRNG in COMA1 is based on AES-GCM and AES-CTR respectively. The COMA2 is implemented by using ACORN for AEAD and Trivium for PRNG, The details of these two variants are summarized in Table \ref{COMA_configss}. The breakdown of area (in terms of Slices, LUTs, and FFs) for these solutions for an FPGA implementation in Xilinx Artix-7 (using Minerva \cite{farahmand2017minerva}) is reported in Table~\ref{FPGA_overhead}. The breakdown of area (in terms of Cells and $um^2$), critical path, and power consumption for an ASIC implementation is reported in Table~\ref{ASIC_overhead}. Note that the 2.5D-COMA needs both the trusted and untrusted parts of the architecture, while the R-COMA only requires the untrusted part. Table \ref{FPGA_top_results} reports optimized area and frequency results on FPGA for top-level of trusted and untrusted sides. As illustrated, the total area of lightweight solution is around 1/3 of the NIST-compliant solution. The reported numbers in Table \ref{FPGA_overhead} include the overhead of all sub-modules including AEAD, CSN-RCSN, RNG, ECC, etc. Due to the optimization on the boundaries among the units, resource utilization in Tables \ref{FPGA_top_results} is less than the sum of row values in Table \ref{FPGA_overhead}.

\begin{table}
\scriptsize   
\centering
\caption{Resource Utilization for ASIC implementation of NIST-compliant and lightweight COMA. \vspace{-2mm}}
\label{ASIC_overhead}
\setlength\tabcolsep{3pt} % default value: 6pt
% \begin{tabular}{|c|c|c|c|c|c|c|c|}

\scalebox{0.9}{

\begin{tabular}{@{} l *4c | *4c @{}}
\toprule 
         {Name} & \multicolumn{4}{c}{AES-GCM+AES-CTR} &\multicolumn{4}{c}{ACORN+Trivium} \\ 
        \cmidrule(lr){2-5}
        \cmidrule(lr){6-9}
                            & Cells     & Area$_{um^2}$ & Tclk$_{ns}$  & Power$_{mW}$   & Cells & Area$_{um^2}$ & Tclk$_{ns}$   & Power$_{mW}$  \\ \midrule
\textbf{COMA}                        & 25338     & 0.11           & 1.97          & 1.62          & 8681  & 0.046         & 1.18          & 0.84          \\ \midrule \midrule
$\triangleright$ RNG       & 5684      & 0.025         & 1.43          & 0.431         & 1267  & 0.007         & 0.27         & 0.144         \\ \midrule
$\triangleright$ CSN/RCSN  & 1749      & 0.008         & 0.08          & 0.11         & 1749  & 0.008         & 0.08         & 0.11         \\ \midrule
$\triangleright$ AEAD      & 13675     & 0.061         & 1.67          & 0.704         & 2257  & 0.013         & 0.97         & 0.251          \\ \midrule
$\triangleright$ ECC      & 3278     & 0.016         & 1.34          & 0.321        & 3278     & 0.016         & 1.34          & 0.321         \\ \midrule
\end{tabular}\vspace{-10pt}
}
\vspace{-8pt}
\begin{flushleft}
\textit{Using Synopsys generic 32nm libraries.}
\end{flushleft}
\vspace{-5pt}
\end{table}

%\vspace{-6pt}
\subsection{COMA Performance} \label{COMA_Perf_DCC}
\vspace{-2pt}

\begin{figure}[t]
\centering
\includegraphics[width = \columnwidth]{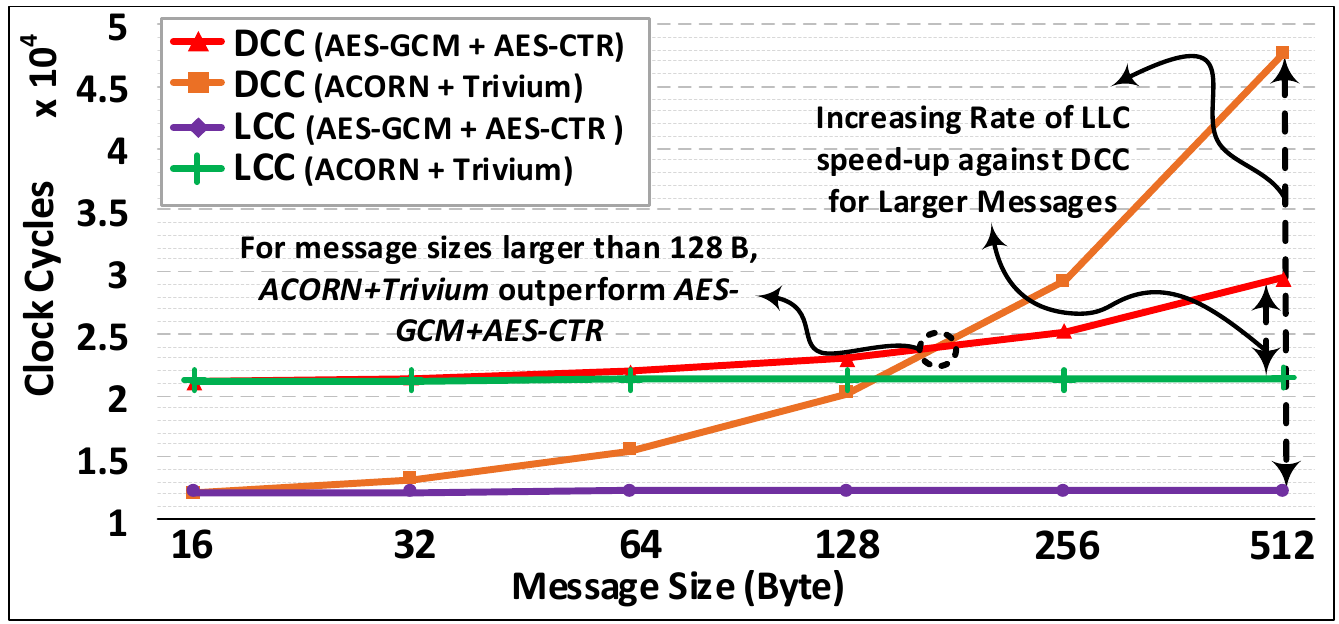}
\caption{Total execution time in number of clock cycles for (AES-GCM $+$ AES-CTR) and (ACORN $+$ Trivium). \vspace{-3mm}}
\label{clk_cycl}
\end{figure}

Fig. \ref{clk_cycl} compares the performance of two solutions in DCC and LCC mode. As illustrated, for small data sizes, the COMA1 outperforms the COMA2 solution. However, as the size of data increases, the COMA2 outperforms the COMA1 solution. It is due to the fact that stream ciphers such as ACORN have a long initialization phase, making them inefficient for small data size. In addition, our AES-GCM implementation benefits from an 8-bit datapath, but the ACORN is realized by a 1-bit serial implementation. The total latency in terms of the number of clock cycles for COMA1 and COMA2 implementations can be calculated using equation (\ref{delay_calc_formula}), in which the number of cycles for the initialization and finalization is fixed and is given in Table~\ref{COMA_configss}. The $C_{byte}$ is the number of cycles needed for encrypting each input message byte, which is 17 and 72 for COMA2 and COMA1, respectively. Hence, in spite of longer initialization, the COMA2 outperforms the COMA1 for message sizes larger than 128 Bytes.    %\vspace{-8pt}
\vspace{-2mm}
\begin{equation}
       T_{comm} = C_{fix} +Message_{size}\times C_{byte}
\label{delay_calc_formula} 
\end{equation}
\vspace{-4mm}

\begin{table}
\scriptsize 
\centering
\caption{Optimized results of COMA Architecture for NIST-compliant and lightweight solution.}
\label{FPGA_top_results}
\setlength\tabcolsep{4pt} % default value: 6pt
% \begin{tabular}{|c|c|c|c|c|c|c|c|}
\scalebox{0.9}{

\begin{tabular}{@{} l *9c @{}}
\toprule 
         {Name} & \multicolumn{4}{c}{AES-GCM+AES-CTR} & &\multicolumn{4}{c}{ACORN+Trivium} \\ 
        \cmidrule(lr){2-5}
        \cmidrule(lr){7-10}
                & Slice & LUT & FF  & Freq[MHz]   &  & Slice & LUT & FF  & Freq[MHz] \\ \midrule 
Trusted          &     2,297     &   7,094     &  5,892    &     103       &  &     1,030      &  2,901     &    1,924   &   121 \\ \midrule
Untrusted          &   2,818    &  8,781     &   7,169   & 109    &  & 1451  &  4,182    &  3,156 & 120 \\ \bottomrule
\end{tabular}
}
% \begin{flushleft}
% %\textit{* Ratio = \large{($\frac{AES\_GCM+AES}{ACORN+Trivium}$)} }
% \end{flushleft}
\begin{flushleft}
%\vspace{-6pt}
\textit{On Xilinx Artix-7 (XC7A100T-1CSG324) FPGA.}
\end{flushleft}
%\vspace{-13pt}
\end{table}
% %\vspace{-10mm}

% \begin{figure*}[t]
% \centering
% \includegraphics[width = 510pt]{timeline.pdf}
% %\vspace{-6pt}
% \caption{}
% \label{3d_detail_arch}
% \end{figure*}

%\vspace{-10pt}
\subsubsection{COMA performance in LCC mode} \label{COMA_Perf_LLC}
\vspace{-2mm}

In the LCC mode, the AEAD is used to synchronize the initial seed of the PRNG, while the CSN is used for encrypting data. The random (TRN) configuration key for the CSN-RCSN is generated by PRNG, which is updated after transferring every $U$ messages. In COMA, the PRNG has a limited buffer size, and as soon as the buffer is filled with random data, the PRNG stops producing additional bits. The consumption of TRNG output is synchronized (every $U$ messages) and the generation of random inputs is limited to the size of buffer. Hence, the PRNGs in the trusted and untrusted sides are always in sync. The number of cycles it takes to initialize the LCC mode includes the time to initialize the secret key engine ($C_{fix}$), the encryption and transfer and decryption of PRNG seed ($C_{ENC}$), and the time for the PRNG to generate enough output from a newly received TRN ($C_{PRNG}$): %\vspace{-10pt}
%PRNG seed ($C_{ENC}$) of size $Seed_{size}$

\vspace{-2mm}
\begin{equation}\label{cycle_whole_CSN}
    C_{LCC-init} = C_{fix} + C_{ENC} + C_{PRNG}
\end{equation}

Depending on the AEAD used for transferring the original seed, the $C_{fix}$ is obtained from Table \ref{COMA_configss}. The seed size in our implementation is 16 Bytes, hence the $C_{ENC}$ is simply $C_{bytes} \times 16$, and the $C_{PRNG}$ is: 

%\vspace{-4mm}
\begin{equation}\label{c_prng}
   C_{PRNG} =  \frac{Bits_{needed}}{PRNG_{perf}} = \frac{3n \times (log_2n-1)}{PRNG_{perf}}
\end{equation}
%\vspace{-2mm}

Finally, after initialization, and by using a CSN of size $n$ when the bus width of COMA is $BW$, the number of cycles to encrypt and transfer one byte of information is: %\vspace{-8pt}

\begin{equation}\label{cycle_data_CSN}
    C_{byte}^{LLC} = \frac{8}{n} \times (\frac{n}{BW} +1)
\end{equation}
%\vspace{-10pt}

Using a 64-bit CSN and BW of 8 bits, the $C_{byte}^{LLC}=9/8$. Compared to $C_{byte}^{DCC}$ for the COMA1 ($C_{byte}^{DCC}$ =72), and for the COMA2 ($C_{byte}^{DCC}$=17), the LCC mode is at least an order of magnitude faster.  Fig. \ref{clk_cycl} compares the superior performance of LCC mode compare with DCC mode in both COMA variants.

\begin{table*}
\scriptsize 
\centering
%%\vspace{-7mm}
\caption{SAT Execution Time on \emph{OMEGA}-based Blocking CSN and $LOG_{n, log_2(n) - 2, 1}$ as a Close to Non-blocking CSN . \vspace{-2mm}}
\label{sat_time}
\setlength\tabcolsep{6.3pt} % default value: 6pt
\scalebox{0.9}{

\begin{tabular}{@{} l | cc | cc | cc | cc | cc | cc | cc | cc  cc @{}}
\toprule 
CSN Size  & \multicolumn{2}{c}{4} & \multicolumn{2}{c}{8} & \multicolumn{2}{c}{16} & \multicolumn{2}{c}{32} & \multicolumn{2}{c}{64} & \multicolumn{2}{c}{128} & \multicolumn{2}{c}{256} & \multicolumn{2}{c}{512}\\
        \cmidrule(lr){2-3} 
        \cmidrule(lr){4-5}
        \cmidrule(lr){6-7}
        \cmidrule(lr){8-9}
        \cmidrule(lr){10-11}
        \cmidrule(lr){12-13}
        \cmidrule(lr){14-15}
        \cmidrule(lr){16-17}

Mode       	& 	blk	& 	non-blk	& 	blk	& 	non-blk & blk	& 	non-blk		& blk	& 	non-blk		& blk	& 	non-blk		& blk	& 	non-blk		& blk	& 	non-blk		& blk	& 	non-blk		\\ \midrule
SAT Iterations  	& 	6	& 	14	& 	7	& 	18	&   8   &    25	&    12 	& 31  &   14  &  TO & 24 & TO & 25 & TO & TO & TO \\ 
SAT Exe. Time $(s)$  	& 	0.01 & 0.01 & 0.03 & 0.15 & 0.2 & 2.35 & 0.8 & 79.18 & 5.9 & TO & 130.5 & TO & 1136.2 & TO & TO & TO \\ \bottomrule

\end{tabular}
}
\begin{flushleft}
\vspace{-8pt}
\textit{TO: Timeout = $2\times10^6$ seconds; The SAT attack is carried on a Dell PowerEdge R620 equipped with Intel Xeon E5-2670 2.6 GHz and 64GB of RAM. }
\end{flushleft}
\vspace{-3mm}
\end{table*}
\subsubsection{Frequency of TRN updates in LCC mode} \label{TRN_Perf_LCC}
\vspace{-2mm}

The frequency of TRN update ($U$) for LCC is an important design feature.   
A large $U$ reduces energy as PRNG/TRNG is kept idle for $U-P$ cycles. P is the number of required cycles to refill the PRNG buffer after a TRN read. However, when the TRN is fixated for a long duration of time, the possibility of a successful side-channel, algebraic, or SAT attack on the CSN increases. The minimum number of messages required for an algebraic attack (even if such attack is possible) is $n$, which is the CSN input size. Our experiments show that a SAT attack could recover the key with an even smaller number of inputs. Knowing the number of encryptions/decryptions needed by such attacks, we can set the $U$ to a safe value smaller than the number of required messages to make it resistant against these attacks. So, the value of $U$ should be between $P \leq U \leq n$. 

The SAT attack against CSN is implemented similar to \cite{subramanyan2015evaluating}. In this attack the CSN gate-level netlist and an activated chip is available to the attacker, while the attacker aims to extract the CSN-RCSN configuration signals. Table \ref{sat_time} captures the results of the SAT attack against blocking and near non-blocking CSNs. As illustrated, the time to break a near non-blocking CSN is significantly larger. In each iteration SAT test one carefully selected input message. Hence, if the $U$ is kept smaller than the number of required SAT iterations, the SAT attack could not be completed. 

%\vspace{-5pt}
\subsection{Energy saving in LCC mode} \label{TRN_Energy_LCC}
\vspace{-2pt}

\begin{figure}
\centering
% \subfloat[]{{\includegraphics[width=0.8\columnwidth]{ExTruDYN1.pdf} }} \newline %\vspace{-5pt}
  %\vspace{-5pt}
  \begin{subfigure}[b]{\columnwidth}
    \includegraphics[width=0.9\columnwidth]{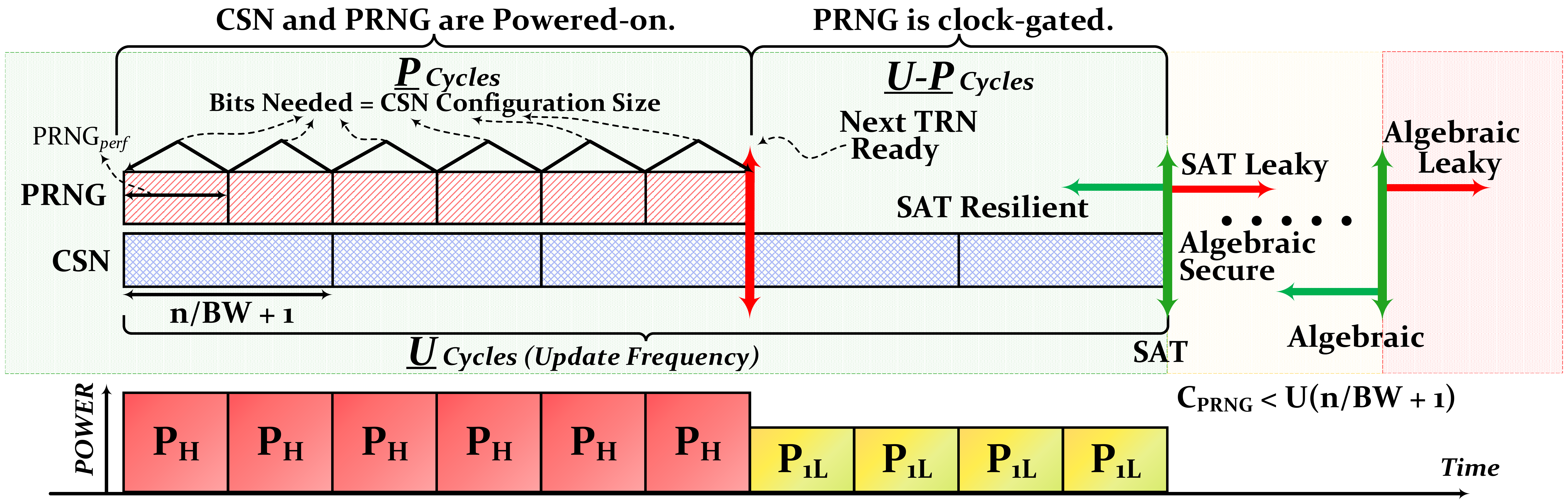}
    \caption{While $P < U$. PRNG is kept idle (power-gated) after $P$ cycles. }
    \label{fig:aesgcmpr}
  \end{subfigure}
  \begin{subfigure}[b]{\columnwidth}
    \includegraphics[width=0.9\columnwidth]{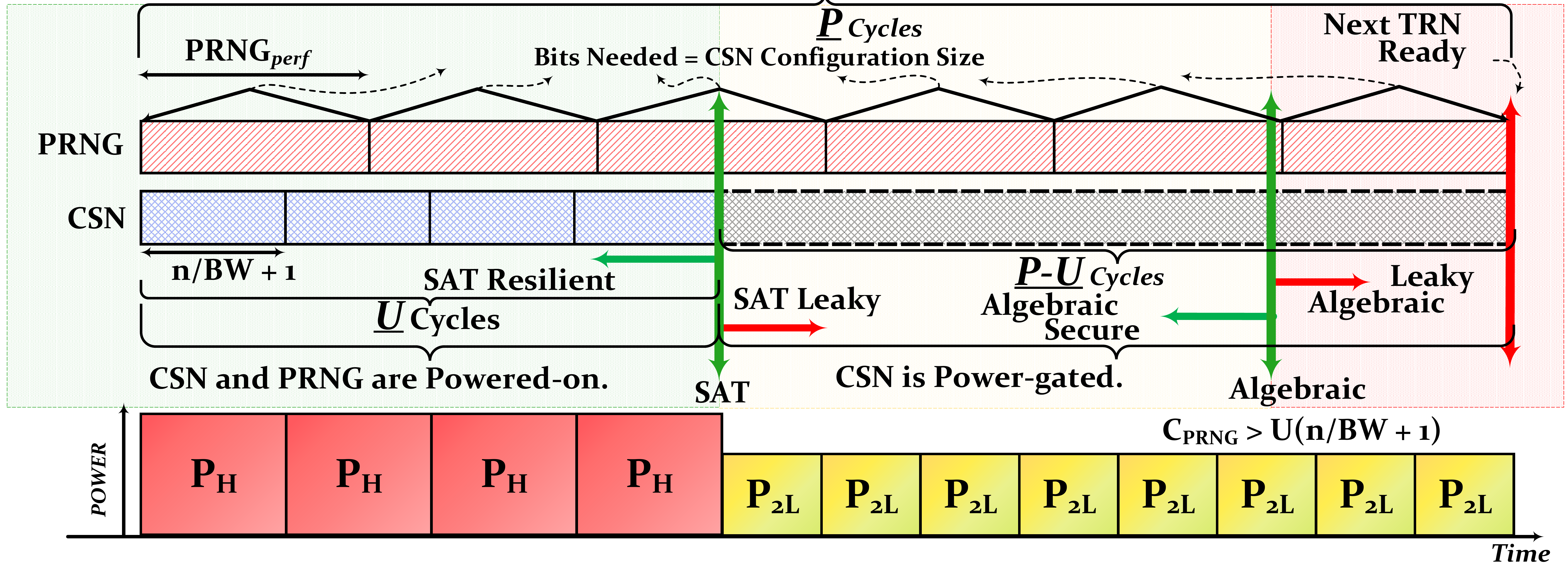}
    \caption{While $U > P$. CSN/RCSN is kept idle after $U$ cycles. \\ \footnotesize{$P_H$ = Power$_{CSN/RCSN}$ + Power$_{PRNG}$;~~~~~$P_{1L}$ = Power$_{CSN}$; $P_{2L}$ = power$_{PRNG}$;}}
    \label{fig:acornunpr}
  \end{subfigure} 
\caption{The Power Consumption at LCC mode of operation.\vspace{-2mm}}
\label{energy_llc}
\vspace{-4mm}
\end{figure}

As illustrated in Fig. \ref{energy_llc}(a), in the LCC mode, the TRN is updated every $U$ cycles. $U$ is determined based on the fastest attack on CSN-RCSN pair, which is the SAT attack. After each TRN update, the PRNG takes $P$ cycles to refill its buffer. Note that $P$ cycles required for PRNG could be stacked at the beginning of $U$ cycles, or distributed over $U$ cycles depending on the size of PRNG buffer. As long as the TRN completely changes every $U$ cycle, the possibility of attack is eliminated. Hence in each $U$ cycles, for $P$ cycles the PRNG/TRNG and CSN are active, and for $U-P$ cycles, the PRNG is clock gated, and only CSN is active. In both cases, the AEAD is active only for the initial exchange of PRNG seed, allowing us to express the power consumption of the LCC mode as: %\vspace{-15pt}
 
\vspace{-4mm}
\begin{equation}\label{energy_llc_eq}
   E_{LLC} =  C_{PRNG}\times P_{H} + \big{(}U(\frac{n}{BW} + 1) - C_{PRNG}\big{)}\times P_{L}
\end{equation}
\vspace{-4mm}

Obviously, the number of required cycles to refill the PRNG buffer after TRN read ($P$) affects energy consumption and communication throughput. If $P < U$, as illustrated in Fig. \ref{energy_llc}(a), for $U-P$ cycles the PRNG is kept idle (power-gated). However, if $P > U$, as shown in Fig. \ref{energy_llc}(b), the communication should be stopped for $P-U$ cycles till the next TRN is ready and to resist SAT or algebraic attacks.

\begin{table*}
\scriptsize
\centering
%\vspace{-3mm}
\caption{COMA vs. FORTIS. \vspace{-3mm} }
\label{comavsfortis}
\setlength\tabcolsep{5pt} % default value: 6pt
\scalebox{0.9}{

\begin{tabular}{@{} l *{9}c @{}}
\toprule
\multicolumn{1}{l}{Scheme} & Key Management  &   Data Comm  &   Private Key   & SC Protected & Session Key & Activation & Need to TPM & RNG\\
\midrule \midrule
FORTIS & Constant & \xmark* & Embedded (known to the fab) & \xmark & Vulnerable to Fault Attack & Once & at Untrusted & PRNG\\
COMA & PUF-based Unique &  \checkmark$^{+}$ & No private key at untrusted & \checkmark & Secure & per Demand & at Trusted & TRNG\\

\bottomrule
\multicolumn{9}{l}{\textit{\textbf{*}: Not Implemented, but Naturally available using OTP. Limited Performance Due to Lightweight RSA}} \\
\multicolumn{9}{l}{\textit{\textbf{$^{+}$}: Available in Two Variant: DCC (Fully Secure and Limited Performance) and LCC (Leaky yet Secure and High Performance).}}
\end{tabular}
}
\vspace{-4mm}
\end{table*}
% %\vspace{-8mm}

% on the other hand is more suited for vast amount of less sensitive data, in which the throughput is crucial. In this case, no interrupt will happen even after each $U$ cycles. The chips continue sending/receiving data to maximize the throughput. However, the communication is leaky and vulnerable to the fast attacks such as algebraic and SAT. As shown in \ref{energy_llc}(b), in option (1) which guarantees the security, the power consumption is lower due to IDLE (interrupted) intervals after each $U$ cycles. But, in option (2) which benefits from maximum throughput, the power consumption is not costly.}

% is more suited for highly sensitive data, in which case since sending data after $U$ cycles may be leaked,

% \begin{figure}
%     \centering
%     \includegraphics[width=\columnwidth]{images/LCC_time_a.pdf}
%     \caption{The Power Consumption at LCC mode of operation. After each U cycles, the TRN input to CSN is updated, which require the PRNG to refill its buffer in the next P cycles. The U is determined by min attack time required to break the CSN-RCSN (determined by SAT attack) and P is determined based on the performance of PRNG.}
%     \label{energy_llc}
% \end{figure}

\begin{figure}
\centering
% %\vspace{-8pt}
\includegraphics[width = \columnwidth]{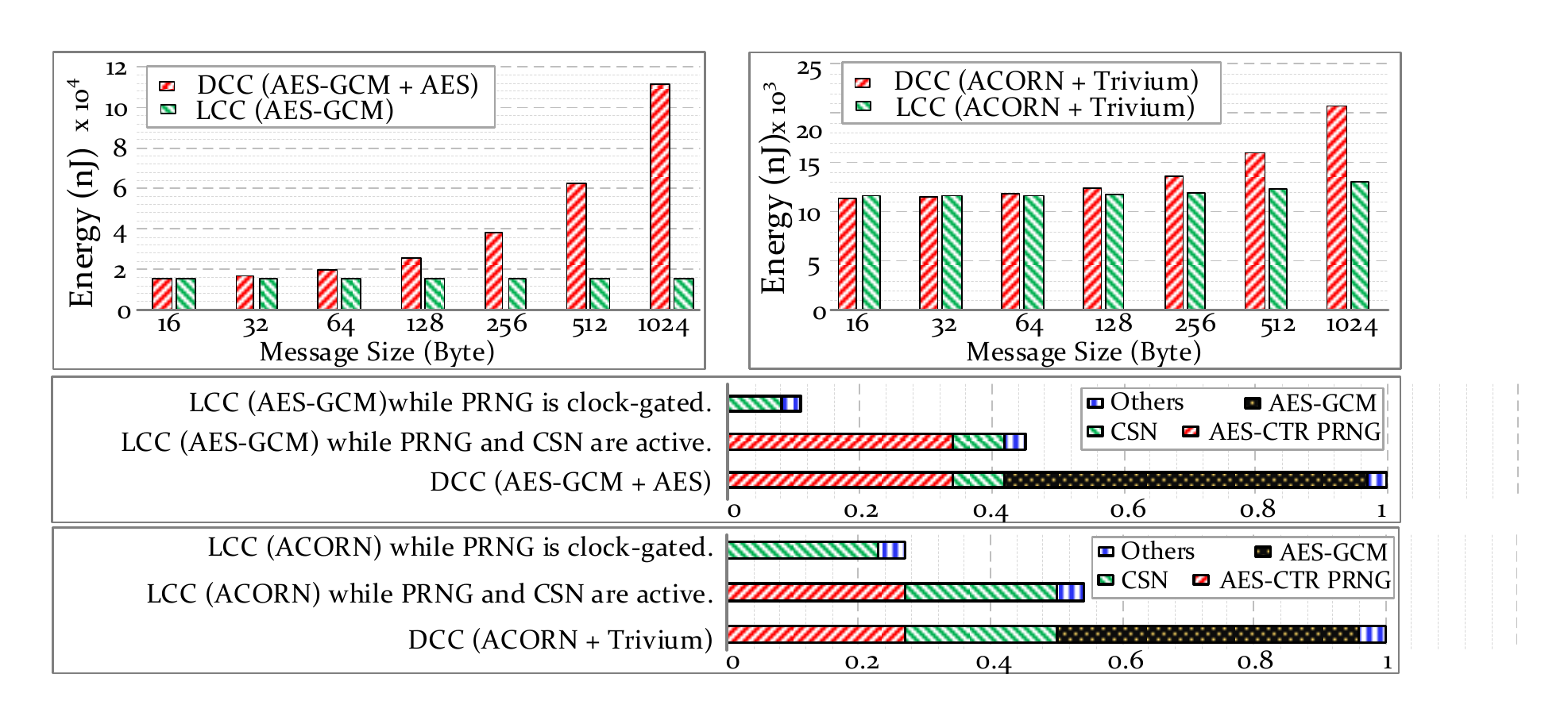}
\vspace{-19pt}
\caption{Energy Breakdown in COMA. }
\label{energy_figures}
\end{figure}

The energy consumption of LCC mode for COMA architectures constructed using NIST-compliant and lightweight solution when transmitting different size of messages is captured in Fig. \ref{energy_figures}.  As illustrated, the LCC mode, for having to synchronize the two sides using a TRNG seed, is burdened with the initialization cost of AEAD. However, when the CSN-RCSN and PRNG are setup, the energy consumed for exchanging additional messages grow at a much lower rate compare to DCC mode (which is dominated by AEAD and PRNG power consumption as reported in table \ref{ASIC_overhead}).

%\vspace{-8pt}
\section{Comparing COMA with Prior Work} \label{comparison_with_prior_work}
\vspace{-2mm}

To the best of our knowledge, FORTIS \cite{guin2016fortis} is the only comprehensive key-management scheme that was previously proposed.
Table~\ref{comavsfortis} compares our proposed solution against FORTIS. COMA addresses several shortcomings of the FORTIS: 

1) In FORTIS, all chips use identical keys, hence there is no mean of differentiating between chips. In COMA each chip has a unique key generated by PUF.  2) In COMA, secret key for communication and authentication is generated by PUF, when FORTIS relies on embedding the private key and public key in GDSII. So, the private key in FORTIS will be known to the fabrication posing the risk that the entire process of activation could be faked in software. In COMA, such attack is prevented as secret key is generated by PUF and is securely read out using public key cryptography. 3) In FORTIS, the usage of the private key for chip authentication is vulnerable to SCA. In COMA, the secret-key cryptography is side channel protected, and the public-key encryption is only used once, making COMA secure against SCA. 4) In FORTIS, there is also the possibility of deploying a fault attack by fixing the value of session key Ks. In COMA, the same attack would require fixing the PUF output or replacing the PUF with a known function. This however could be tested by reading out the output of the PUF using multiple challenges and performing statistical test on the PUF response (PUF health check). 5) In FORTIS, the activation is done once, hence there is a need to store the obfuscation key in the untrusted chip. In COMA, the need to store the obfuscation key in untrusted chip is removed. In R-COMA, the activation takes place on demand, and the key is removed after power down or reset. In 2.5D-COMA, the activation key is stored in a trusted chip. 6) COMA provides two new mechanisms for communication: a) the DCC mode for added security, and b) the LCC mode for high-speed communication. 7) COMA uses a TRNG to produce the seed for PRNG, while FORTIS uses a PRNG without addressing a random source for its seed, increasing its vulnerability. 

In terms of area overhead, FORTIS \cite{guin2016fortis} provides an estimate for the incurred overhead of their solution, which is around 10K gates. As shown in Table \ref{ASIC_overhead}, the numbers of cells for implementing the NIST-compliant (COMA1) implementation is 25.4K gates, while the lightweigh solution (COMA2) is implemented using 8.7K gates. Table~\ref{benchs} compares the area overhead of FORTIS against COMA1 and COMA2, when these architectures are deployed to protect a few mid- and large-size benchmarks. Using COMA2, which improves the overhead by 14\% compared to FORTIS, requires between 0.43\% and 21.3\% of circuit area in selected benchmarks.

% \begin{table}[t]
% \footnotesize
% \centering
% \caption{Comparison of Different Trust Approaches. %\vspace{-3mm}}
% \label{comacompare}
% \setlength\tabcolsep{3.5pt} % default value: 6pt
% \begin{tabular}{@{} l *{9}c @{}}
% \toprule
% Scheme & IC Overproduction & SC Protection & Unique Key per Chip \\
% \midrule \midrule
% Logic Obfuscation & \checkmark & N/A & \xmark \\
% HW Watermarking & \xmark & N/A & \xmark \\
% IC Metering & \xmark & N/A & \xmark \\
% FORTIS & \checkmark & \xmark & \xmark \\
% COMA & \checkmark & \checkmark & \checkmark \\
% \bottomrule
% \end{tabular}
% %\vspace{-8pt}
% \end{table}

%%\vspace{-2mm}
\begin{table}[t]
\scriptsize
\centering
%\vspace{-2mm}
\caption{Area Overhead of COMA vs. FORTIS. \vspace{-2mm}}
\label{benchs}
\setlength\tabcolsep{2.5pt} % default value: 6pt
\scalebox{0.9}{

\begin{tabular}{@{} l  *5c @{}}
\toprule
\multicolumn{1}{l}{Design} & Gate Count & FORTIS/Design & COMA1/Design & COMA2/Design\\
\midrule
\midrule
b19    & 40,789 & 24.52\% & 62.1\% & 21.28\% \\
\midrule
VGA\_LCD & 43,346 & 23.07\% &  58.45\% & 20.02\% \\
\midrule
Leon3MP & 253,050 & 3.95\% & 10.01\% & 3.43\% \\
\midrule
SPARC & 836,865 & 1.19\% & 3.02\% & 1.03\% \\
\midrule
Virtex-7 & 2M & 0.5\% & 1.26\% & 0.43\% \\  
\bottomrule
\end{tabular}
}
\vspace{-4mm}
\end{table}
% %\vspace{-5pt}

% \textcolor{magenta}{In FORTIS, it is assumed that usage of DFT compression engines would eliminate the risk of attacks (sensitization and possibly SAT) attack due to inability of the attacker in reading out the actual register values in an uncompressed format. We, however, believe that the claimed resistance of the DFT compression engine has to be investigated. Hence, in our work, even in the presence of DFT compression hardware, we assume that a SAT resilient obfuscation scheme such as \textcolor{red}{[FLL ref]} is employed.} 

\vspace{-2mm}
\section*{Acknowledgement}
\vspace{-2mm}

This research is funded by the Defense Advanced Research Projects Agency (DARPA \#FA8650-18-1-7819) of the USA, and partly by Silicon Research Co. (SRC TaskID 2772.001) and National Science Foundation (NSF Award\# 1718434).

\vspace{-2mm}
\section{Conclusion}  \label{conclusion}
\vspace{-2mm}

In this paper we presented COMA, an architecture for obfuscation-key management and metered activation of an obfuscated IC that is manufactured in an untrusted foundry, while securing its communication. The proposed solution removes the need to store the key in the untrusted chip, makes the obfuscation unlock-key a moving target, allows unique identification of the protected IC, and secures the communication to/from the protected chip using two hybrid cryptographic schemes for ultra-high-speed and ultra-security. Our experimental results show that compared to the state-of-the-art key management architecture, FORTIS, COMA is able to reduce the area overhead by 14\%, while addressing many of the shortcomings of the previous work.

%-------------------------------------------------------------------------------
%%%%\vspace{-2mm}
\bibliographystyle{plain}
\bibliography{refs}

%%%%%%%%%%%%%%%%%%%%%%%%%%%%%%%%%%%%%%%%%%%%%%%%%%%%%%%%%%%%%%%%%%%%%%%%%%%%%%%%
\end{document}